\documentclass[twocolumn]{aastex63}

\pdfoutput=1

\usepackage{natbib}
\usepackage{color}
\usepackage{bm}
\usepackage{hyperref}
\usepackage{float}
\usepackage{latexsym}
\usepackage{amsmath}
\usepackage{amssymb}
\usepackage{multirow}
\usepackage{rotate}
\usepackage{graphicx}
\usepackage{url}
\usepackage{color}
\usepackage[hyphens]{url}

\submitjournal{ApJ}
\received{\today}

\begin{document}

\title{White Dwarfs in the Era of the LSST and its Synergies with Space-Based Missions}

\shorttitle{White Dwarfs in the 2020s}
\shortauthors{Fantin et al.}

\smallskip

\author[0000-0003-3816-3254]{Nicholas J. Fantin}
\correspondingauthor{Nicholas J. Fantin}
\email{nfantin@uvic.ca}
\affil{Department of Physics and Astronomy, University of Victoria, Victoria, BC, V8P 1A1, Canada}
\affil{National Research Council of Canada, Herzberg Astronomy \& Astrophysics Research Centre, 5071 W. Saanich Rd, Victoria, BC, V9E 2E7, Canada}

\author{Patrick C{\^o}t{\'e}}
\affil{National Research Council of Canada, Herzberg Astronomy \& Astrophysics Research Centre, 5071 W. Saanich Rd, Victoria, BC, V9E 2E7, Canada}

\author{Alan W. McConnachie}
\affil{National Research Council of Canada, Herzberg Astronomy \& Astrophysics Research Centre, 5071 W. Saanich Rd, Victoria, BC, V9E 2E7, Canada}

\begin{abstract}

With the imminent start of the Legacy Survey for Space and Time (LSST) on the Vera C. Rubin Observatory, and several new space telescopes expected to begin operations later in this decade, both time domain and wide-field astronomy are on the threshold of a new era. In this paper, we use a new, multi-component model for the distribution of white dwarfs (WDs) in our Galaxy to simulate the WD populations in four upcoming wide-field surveys (i.e., LSST, Euclid, the Roman Space Telescope and CASTOR) and use the resulting samples to explore some representative WD science cases.  Our results confirm that LSST will provide a wealth of information for Galactic WDs, detecting more than 150 million WDs at the final depth of its stacked, 10-year survey. Within this sample, nearly 300,000 objects will have 5$\sigma$ parallax measurements and nearly 7 million will have 5$\sigma$ proper motion measurements, allowing the detection of the turn-off in the halo WD luminosity function and the discovery of more than 200,000 ZZ Ceti stars. The wide wavelength coverage that will be possible by combining LSST data with observations from Euclid, and/or the Roman Space Telescope, will also discover WDs with debris disks, highlighting the advantages of combining data between the ground- and space-based missions.

\bigskip

\end{abstract}

\section{Introduction}
\label{sec:intro}

The upcoming decade will see the construction of several powerful new observatories with wide-ranging science cases. Previous generation sky surveys, from the Palomar Observatory Sky Survey \citep{POSS, POSS2}, to more recent surveys like the Sloan Digital Sky Survey \citep{York2000} and \textit{Gaia} \citep{Gaia}, have established that an incredible breadth of research can be undertaken with publicly-accessible, wide-field survey data. This includes the study of white dwarfs (WDs) --- the end stage of stellar evolution for the vast majority of stars.

WDs have been used, among other things, to test fundamental physics \citep[see, e.g,][]{Winget2008, Hansen2015}, to study stellar populations of all ages \citep[see, e.g,][]{Richer1997,Bedin2009, Salaris2018}, and to explore Milky Way formation and evolution \citep{Winget1987,Rowell2013, Kilic2017, Fantin2019}. This utility is because WDs represent extreme physical environments and yet are ubiquitous throughout the Milky Way. 

Despite their importance, detecting these inherently faint objects can be a challenge. Historically, WDs have been discovered in large numbers in wide-field, multi-epoch, optical surveys as either high proper motion objects \citep[e.g,][]{Luyten1979, RowellHambly2011}, or as star-like objects with a UV excess \cite{Green1986}. Such surveys have uncovered hundreds to thousands of WDs, with the majority residing within the local neighborhood. 

\begin{figure*}[!t]
	\centering
	\includegraphics[angle=0,width=\textwidth]{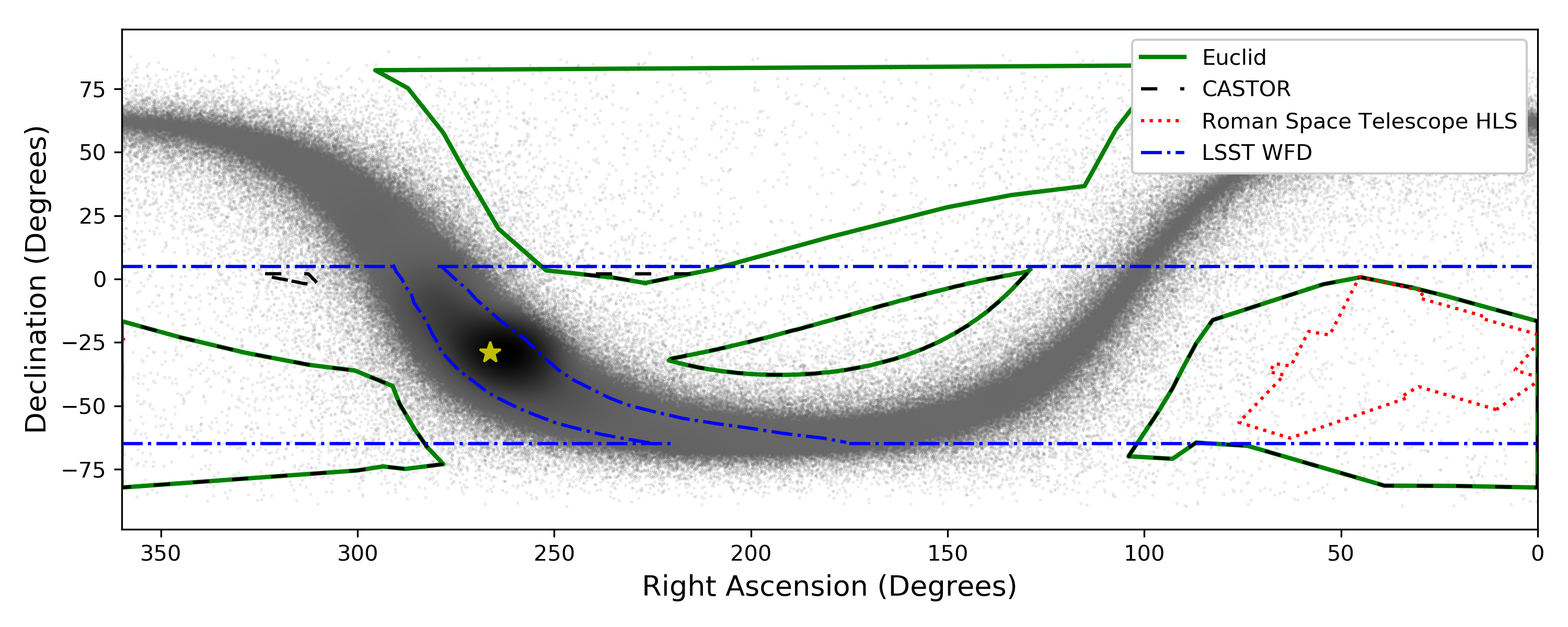}\\
	
	\includegraphics[angle=0,width=\textwidth]{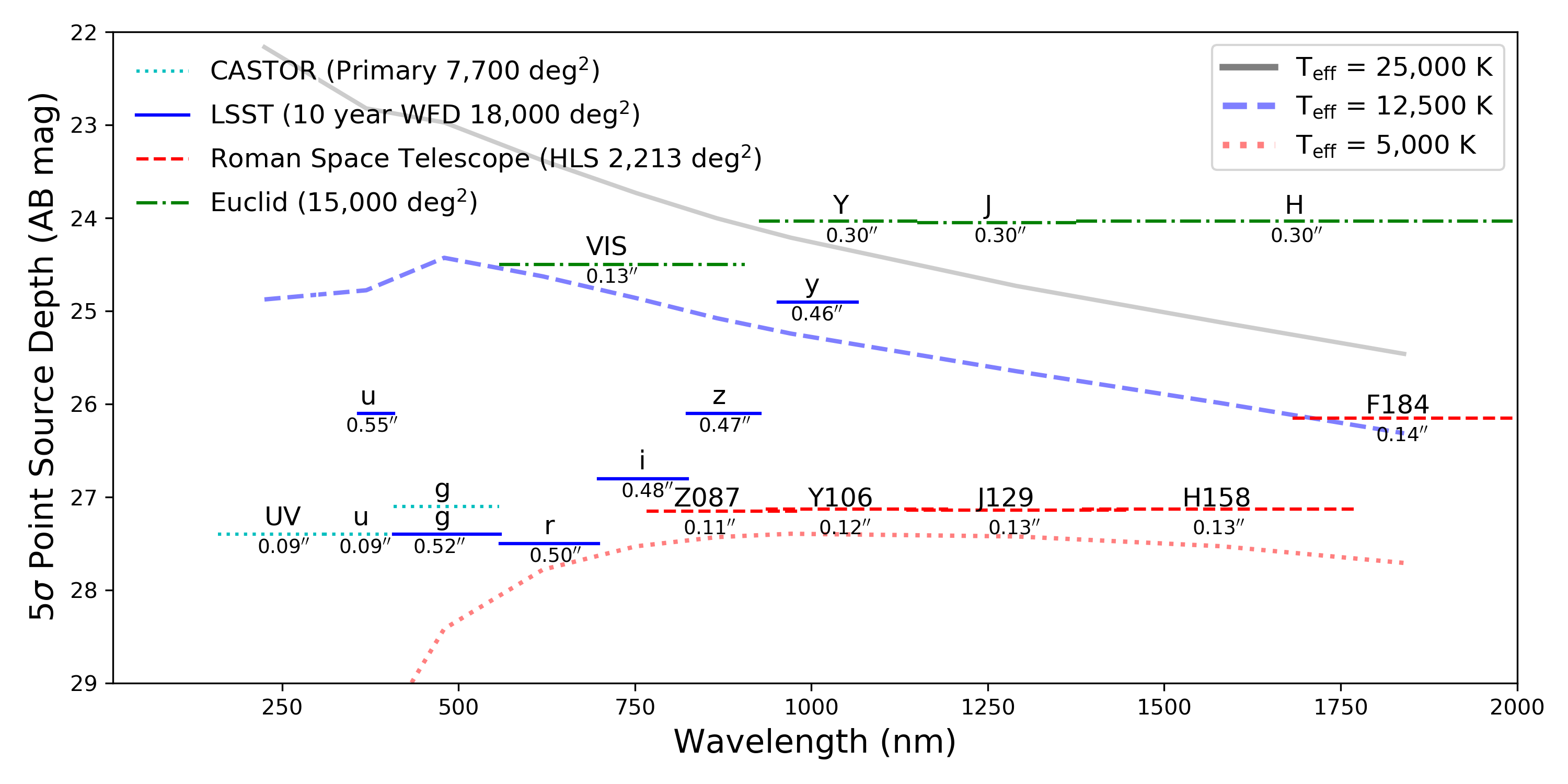}
	
	\caption{\textit{Top:} Footprints for the various surveys in equatorial coordinates. Also plotted are model white dwarfs for reference, ignoring extinction, to highlight the location of the Galactic disk. The Galactic center is marked with a star. \textit{Bottom:} Magnitude limits, wavelength coverage, and median image quality for each survey. Also plotted are the SEDs for 0.6\,M$_{\odot}$ white dwarfs of three temperatures each placed at a distance of 3,800\,pc: 25,000\,K (solid black), 12,500\,K (dashed blue), and 5,000\,K (dotted red) for reference.  \bigskip }
	\label{fig:maglimits}
\end{figure*}

Spectroscopic and photometric samples of WDs increased by orders of magnitude with the beginning of the Sloan Digital Sky Survey \citep[SDSS;][]{York2000, Kleinman2004}. Photometrically, WD samples on the order of $\sim$10$^4$ have been generated using SDSS data as a second epoch to derive proper motions \citep{Harris2006, Munn2017}, allowing for an accurate determination of the local luminosity function. Furthermore, the spectroscopic sample currently stands at more than 20,000 objects as of SDSS Data Release 14 \citep{Kepler2019}, although this sample was acquired mainly as a result of a targeted QSO survey which imparts a significant temperature bias on the sample \citep[see e.g,][]{Kleinman2013}.

WDs can also be identified using parallaxes as their intrinsic faintness allows them to be differentiated from other stars with equivalent colors. With the release of \textit{Gaia} DR2, the number of WD candidates has increased by another order of magnitude, with some catalogs containing on the order of 10$^5$ WDs \citep{JimenezEsteban2018,Gentile2019}. 

Because the magnitude limit of the \textit{Gaia} catalog is only $G\sim$ 20.5, future surveys are poised to further increase the sample of known WDs by many more orders of magnitude. In this paper, we examine the scope of WD science that can be performed using some major ground- and space-based surveys that will be carried out later this decade --- the Legacy Survey for Space and Time (LSST), Euclid, and the Roman Space Telescope. We also consider the proposed Cosmological Advanced Survey Telescope for Optical and uv Research (CASTOR), which would provide UV/blue-optical imaging that is complementary to these other facilities.

We begin by describing the key aspects of each survey in \S\ref{sec:surveys} and then briefly describe our model in \S\ref{sec:model}. The simulated WD samples expected within each survey are discussed in \S\ref{sec:results} before we discuss in \S\ref{sec:selection} the various WD selection options using the LSST. In \S\ref{sec:HLS}, we highlight some notable WD research opportunities that would be possible in the Roman Space Telescope High Latitude Survey (HLS), a $\sim$2000 deg$^2$ region that will be included in all surveys. We finish with an analysis of some other, representative WD science cases in \S\ref{sec:discussion} before summarizing in \S\ref{sec:summary}.

\section{The Surveys}
\label{sec:surveys}

In this section, we describe four planned surveys with nominal start times in this decade. We highlight the relevant capabilities including spatial coverage, photometric depths, and whether any astrometric measurements will be performed.

\subsection{The Legacy Survey for Space and Time (LSST)}
\label{subsec:lsst}

The Simonyi Survey Telescope (formerly the Large Synoptic Survey Telescope), part of the newly named Vera C. Rubin Observatory, is an upcoming ground-based observatory in the Southern Hemisphere. With an 8.4m primary mirror and nearly 10 deg$^2$ field of view, the observatory will efficiently image much of the Southern sky for a period of 10\,years with a focus on time-domain astronomy. The resulting survey, The Legacy Survey of Space and Time, will produce precise photometry and astrometry nearly 4 magnitudes deeper than the current \textit{Gaia} mission over a large swath of the Southern sky \citep{lsst, Ivezic2019}.

While the survey itself consists of a few different regions of the sky, in this paper we focus on the main survey, dubbed the ``wide-fast-deep" (WFD) survey. This survey will cover the declination range $-65 \le \delta \le +$5 while avoiding a region of the Galactic Plane (see the top panel of Figure \ref{fig:maglimits}). The survey will produce a single visit 5$\sigma$ point-source $r$-band magnitude limit of 24.5, culminating with co-added depths of $r\sim$27.5 at the completion of the survey \citep{lsst}.

The survey will be performed in six optical bands, $ugrizy$, as shown in the bottom panel of Figure \ref{fig:maglimits}. The survey will also be performed under very good seeing conditions, with median conditions between 0.55$^{\prime\prime}$ in the $u$-band to 0.46$^{\prime\prime}$ in the $y$-band.

A key strength of the LSST is its ability to measure accurate parallax and proper motion measurements to depths previously only attainable in space-based missions. The survey achieves this by imaging each part of the sky in two 15-sec exposures, called a visit, more than 800 times per source \citep{lsst2}. This will revolutionize the field of time-domain astronomy, which includes everything from the most distant supernovae to more local stellar populations like white dwarfs.

The Observatory is currently being constructed at Cerro Pach{\'o}n in the Andes mountains of Chile, with first light expected in 2021. The main survey is expected to begin in October 2022 and last 10\,years, producing more than 200 Petabytes of data in the process.

\subsection{Euclid}
\label{subsec:euclid}

The Euclid space mission \citep{Euclid} is the European Space Agency's upcoming near-infrared (NIR) survey designed to investigate dark energy through weak lensing and baryonic acoustic oscillation measurements. This will be accomplished by surveying billions of galaxies over $\sim$15,000 deg$^2$ of sky \citep{Racca2016}. However, like many large surveys, there are many other ancillary science cases. Relevant for this study is the ability of NIR imaging for WDs to reveal dusty debris disks and sub-stellar companions like cool M-dwarfs and brown dwarfs \citep{Dennihy2017}.

Euclid's Wide Survey will focus on high Galactic latitude fields, imaging in three NIR bands ($JHK$), and one wide optical band ($VIS$) to 5$\sigma$ depths of $\sim$24 in the NIR and $\sim$25.5 in the optical. With a 1.2m primary mirror and 0.5 deg$^2$ field-of-view, the resulting images will have excellent resolution: i.e., between 0.1$^{\prime\prime}$ and 0.3$^{\prime\prime}$. 

As Figure \ref{fig:maglimits} shows, the main survey field (solid green) will overlap with many of the other surveys, including nearly 40\% of the LSST WFD survey, nearly the entire CASTOR primary survey, and the entire Roman Space Telescope HLS. The launch time as of publication is expected to be in 2022\footnote{\url{https://www.euclid-ec.org/}}, with an expected mission duration of seven years, meaning that all data will be in hand before the completion of the LSST.

\subsection{The Nancy Grace Roman Space Telescope}
\label{subsec:wfirst}

The Nancy Grace Roman Space Telescope, shortened as the Roman Space Telescope, and formerly the Wide-Field InfraRed Survey Telescope (WFIRST) \citep{WFIRST}, is an upcoming NASA mission that also aims to study dark energy. The telescope consists of a 2.4m primary mirror with a field-of-view of 0.28 deg$^2$. The Roman Space Telescope's High Latitude Survey will image roughly 2,200 deg$^2$ of the Southern Sky, in a region that fully enclosed within the LSST WFD footprint. 

The survey will include optical and NIR imaging, from 5000\,\AA~to 2\,$\mu$m to a depth of $>$26.5 AB mag. The image quality will be roughly 0.1$^{\prime\prime}$ (see Figure \ref{fig:maglimits}). Since the imaging will be collected over the 5\,year duration of the mission, absolute proper motions for many stars will be measured \cite{WFIRST}, allowing for many ancillary Galactic science cases, including the detection of some of the coolest WDs in the Galactic halo. The Roman Space Telescope is set to launch in the mid-2020s. If this time-frame is realized, then the Roman Space Telescope's data products will be available before the completion of LSST's 10-year survey.

\subsection{The Cosmological Advanced Survey Telescope for Optical and uv Research (CASTOR)}
\label{subsec:castor}

The Cosmological Advanced Survey Telescope for Optical and uv Research (CASTOR) is being developed by the Canadian Space Agency (CSA) for a possible launch in the late 2020s. This 1m diameter, off-axis telescope would use a three-mirror anastigmat design to deliver high-resolution imaging (FWHM $\simeq$ 0\farcs15) over an instantaneous 0.25 deg$^2$ field of view \citep{Castor, castor2}. Dichroics and multi-layer coatings would be used to define its photometric passbands --- UV (0.15--0.30 $\micron$), u$^\prime$ (0.30--0.40 $\micron$) and g (0.40--0.55 $\micron$) --- and deliver simultaneous images to three distinct focal plane arrays. 

Among its various science programs, CASTOR would carry out a ``primary survey" that would image the $\sim$7700 deg$^2$ region defined by the overlap of the LSST WFD, Euclid Wide and the Roman Space Telescope HLS footprints. This survey would reach a (5$\sigma$) point-source depth of $m_{\rm AB} \sim 27.2$~mag. By focusing on the UV/blue-optical region, CASTOR would complement the Euclid and the Roman Space Telescope missions by expanding the available SED coverage; relative to LSST, CASTOR would add coverage in the UV region and provide space-quality resolution at the shortest wavelengths accessible from the ground.

\begin{figure*}[!t]
	\includegraphics[angle=0,width=0.49\textwidth]{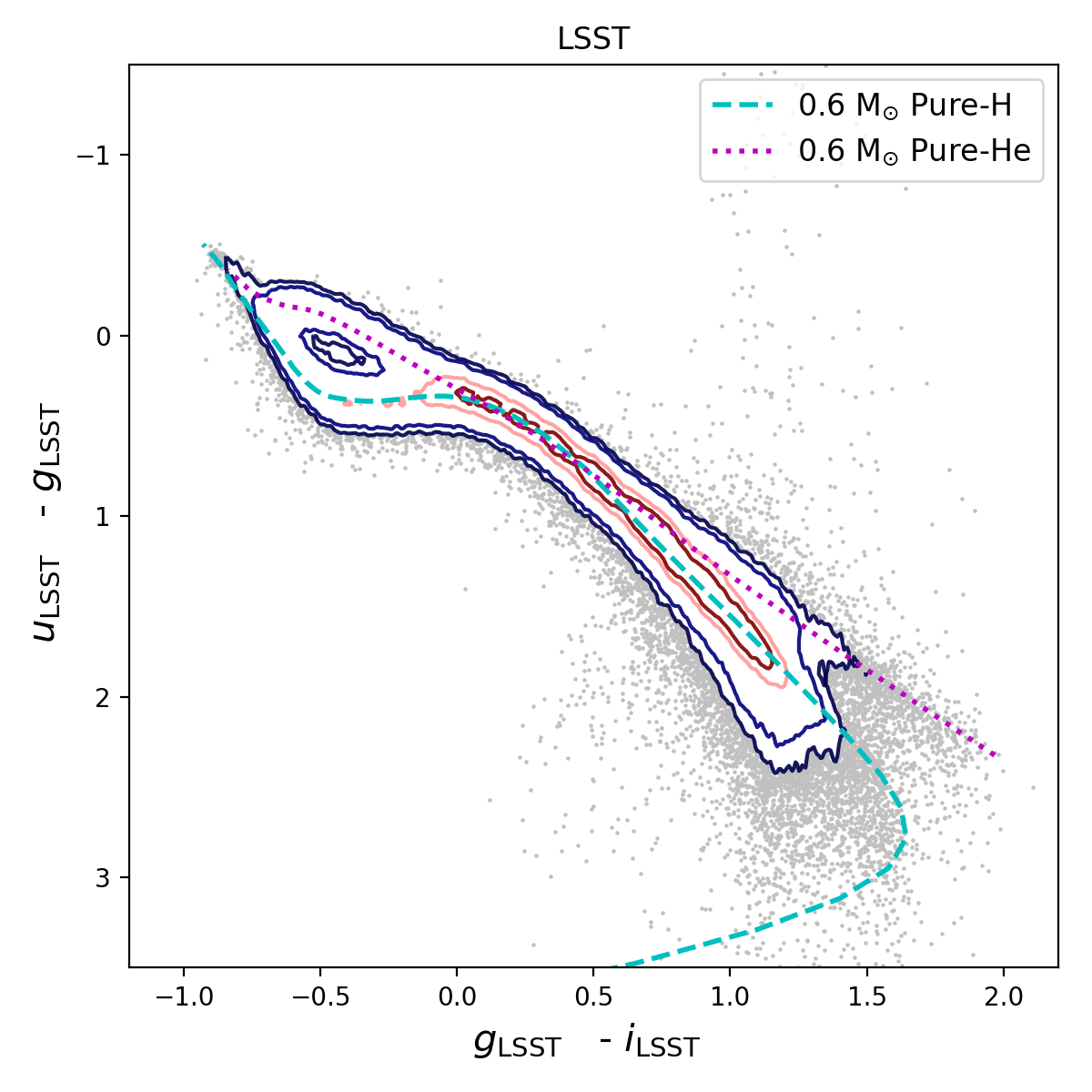}
	\includegraphics[angle=0,width=0.49\textwidth]{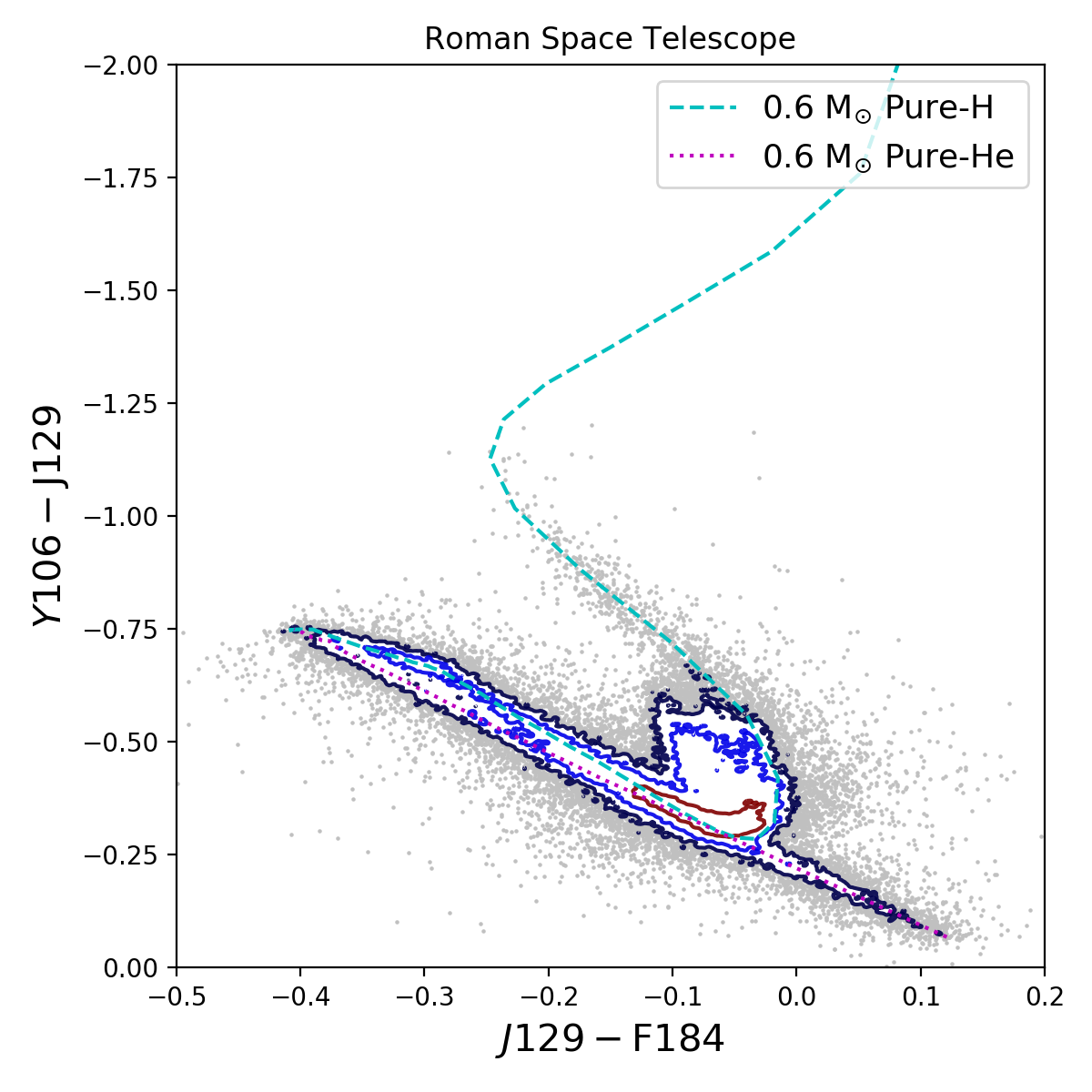}\\
	\includegraphics[angle=0,width=0.49\textwidth]{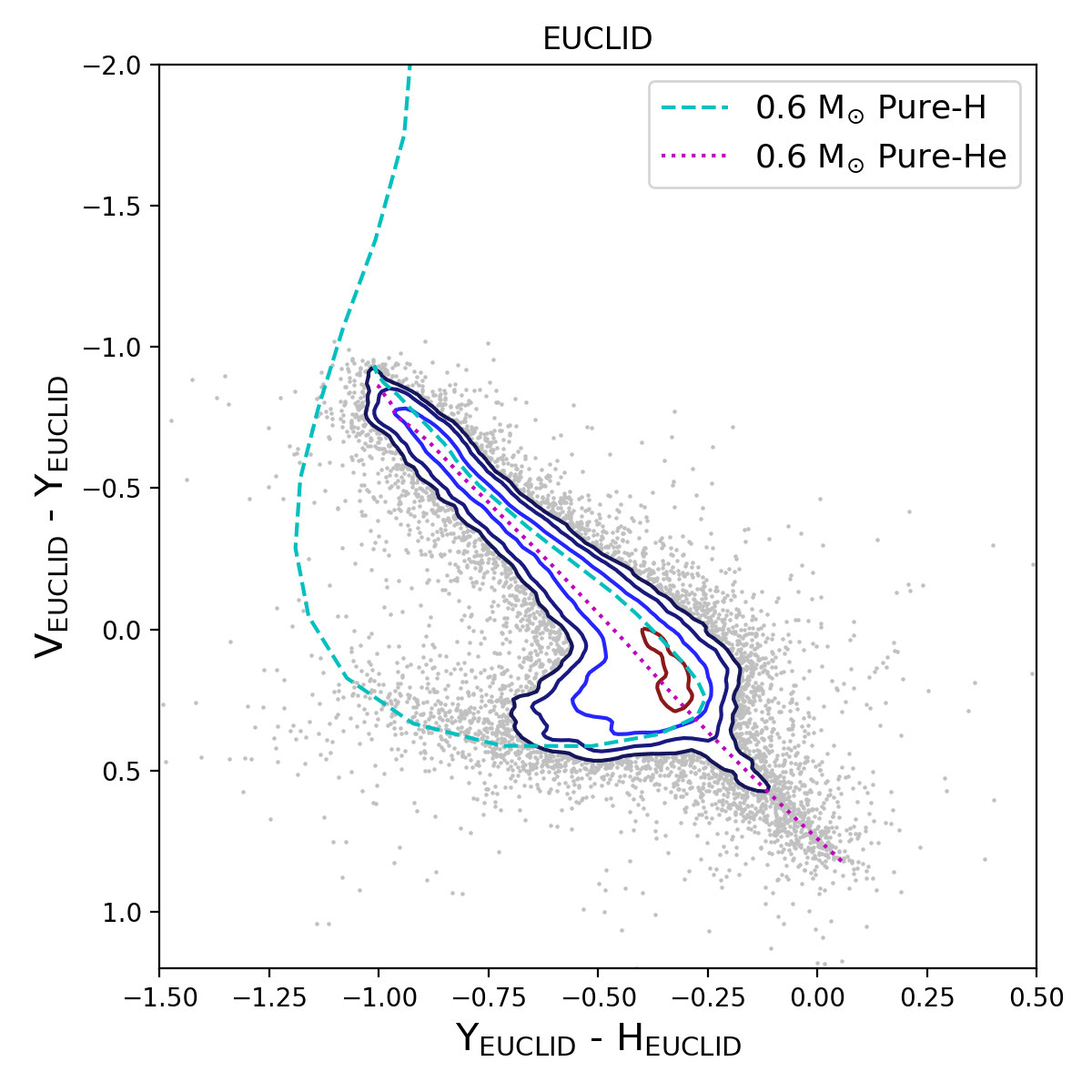}
	\includegraphics[angle=0,width=0.49\textwidth]{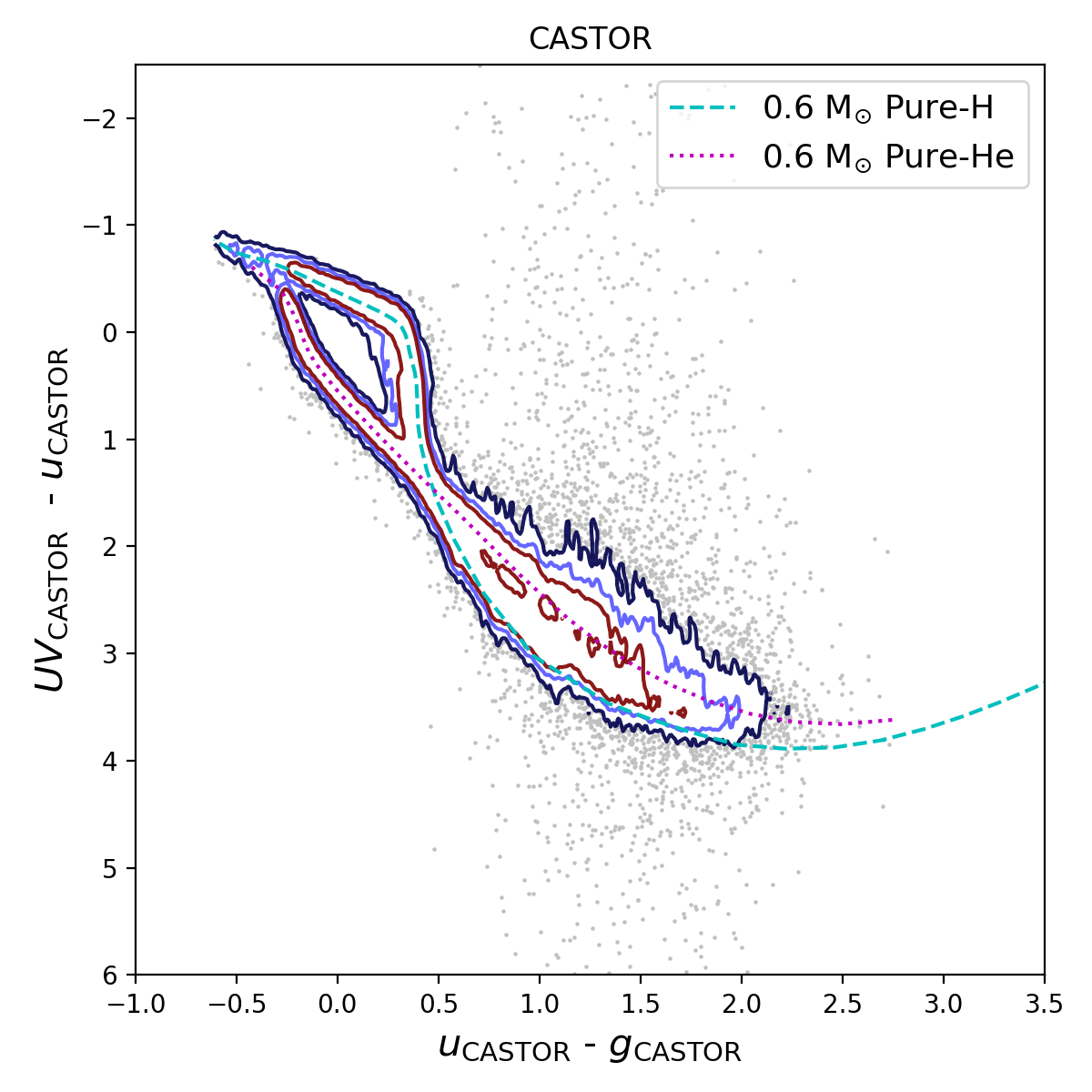}
	\caption{Color-color diagram for the four surveys with model 0.6\,M$_{\odot}$ pure-H (solid cyan) and pure-He (dotted magenta) cooling curves over-plotted. The simulation includes the addition of photometric uncertainties.\bigskip }
	\label{fig:CC}
\end{figure*}

\section{The Model}
\label{sec:model}

We model each survey using the WD population synthesis code presented in \cite{Fantin2019}. This model takes a given star formation history and returns mock catalogs of WDs while taking the survey parameters, selection and completeness effects, as well as the Milky Way geometry, into account. The model consists of three Galactic components --- a thin disk, thick disk, and stellar halo --- each with a prescribed density structure, metallicity, and star formation history. The two disk components are assumed to be double exponential profiles in the radial and vertical direction and evenly distributed in the angular coordinates. The scale heights are set to 300 and 800\,pc for the thin and thick disk respectively. We adopt a constant metallicity for each component equal to [Fe/H] = 0.0 for the thin disk, [Fe/H] = $-$0.7 for the thick disk, and [Fe/H] = $-$1.5 for the halo.

\begin{table}[!t]
    \label{table:passbands}
	\centering
		\caption{Photometric Information}
		\begin{tabular}{lccc}
			\hline
			Passband & $\lambda_{\mathrm{mean}}$($\mathrm{\text{\AA}}$)& A$_{\lambda}$ & Reference \\
			\hline
            lsst u & 3693.2 &  4.145 &  a \\
            lsst g & 4797.3 &  3.237 &  a \\
            lsst r & 6195.8 &  2.273 & a \\
            lsst i & 7515.3 &  1.684 & a \\
            lsst z & 8664.4 &  1.323 & a \\
            lsst y & 9710.3 &  1.088 & a \\
            \hline
            CASTOR UV& 2290.0  & 7.070 &  b \\
            CASTOR u & 3460.0  & 4.249 &  b \\
            CASTOR g & 4800.0  & 3.230 &  b \\
            \hline
            Roman Space Telescope R062&  6200.0  &  2.374 & c \\
            Roman Space Telescope Z087&  8685.0  &  1.436 & c\\
            Roman Space Telescope Y106&  1059.5  &  1.023& c\\
            Roman Space Telescope J129&  1292.5  &  0.755 & c\\
            Roman Space Telescope H158&  1577.0  &  0.541 & c\\
            Roman Space Telescope F184&  1841.5  &  0.413 & c\\
            \hline
            Euclid VIS& 7250.0  & 1.781 &  c\\
            Euclid Y  & 1033.0  & 1.004 & c\\
            Euclid J  & 1259.0  & 0.702 & c\\
            Euclid H  & 1686.0  & 0.455 & c\\

		\end{tabular}
        \tablecomments{We assume R$_{V}$ = 3.1.\\ a -- \citep{Schlafly2011}. \\
        b -- Using Equation A1 of \citep{Schlafly2011}\\
        c -- Padova Isochrones \url{http://stev.oapd.inaf.it/cmd} \\}
		\bigskip
		\bigskip
\end{table}

The star formation history (SFH) for the model was calibrated using photometric data from the Canada France Imaging Survey \citep[CFIS;][]{CFIS1}, Pan-STARRS1 DR1 3$\pi$ \citep[PS1;][]{PS1}, and \textit{Gaia} DR2 \citep{Gaia} in \cite{Fantin2019}. Each component consists of a skewed Gaussian, with a mean, standard deviation, and skew. We apply our resulting star formation history, seen in Figure 10 of \cite{Fantin2019}, to these model instances. 

Stars are spawned with a given mass, sampled from a Kroupa IMF \cite{Kroupa2001}, and at a given position and time based on the spatial distribution and SFH. The pre-WD lifetime is based on the initial mass and is calculated using the functional forms presented in \cite{Hurley2000}. A star will form a WD if its main-sequence age is less than the time between formation and present day, with the difference being the cooling age. We allow the WDs to have either a pure-hydrogen (DA) or pure-helium (DB) atmosphere, with the DB fraction being (21 $\pm$ 3)\,\% as determined in \cite{Fantin2019}.

The WD mass is then calculated using the initial-to-final mass relation of \cite{Cummings2018}. With the mass, cooling age, and atmospheric type we calculate the photometric magnitudes using the cooling models of \cite{HolbergBergeron2006}, which have specifically been computed using the transmission curves for each survey. The WD cooling sequences are similar to those described in \citet{Fontaine2001} with (50/50) C/O-core compositions, $M_{\rm He}/M_{\star}=10^{-2}$, and $M_{\rm H}/M_{\star}=10^{-4}$ or $10^{-10}$ for H- and He-atmosphere WDs, respectively\footnote{See \url{http://www.astro.umontreal.ca/~bergeron/CoolingModels}}. An extinction correction is then applied using the E(B$-$V) values from \cite{Schlafly2011} using the \textsc{DUSTMAPS} python package developed by \cite{DustMaps2}. The photometric extinction coefficients used for each survey are shown in Table \ref{table:passbands}. We do not apply extinction corrections for objects closer than 100\,pc, and instead begin applying extinction linearly between 100 and 250\,pc before applying a full extinction correction beyond 250\,pc as in \cite{Kilic2017}.

We also simulate a mock proper motion by sampling from velocity ellipsoids for each component using results from \cite{Robin2017}. These velocities are converted into a proper motion using the equations from \cite{JohnsonSoderblom1987} with updated values for the Galactic poles.

Finally, we apply the survey parameters, including the survey area and magnitude completeness corrections, as presented in Figure \ref{fig:maglimits}. The completeness functions are applied as in Figure 4 of \cite{Fantin2019} since they allow for a smooth drop-off beyond the 5$\sigma$ limits, and we apply a shift based on the fainter magnitude limits of each survey. 

The resulting mock catalogs contain an object's position, mass, cooling age, distance, proper motion, Galactic space velocities, as well as the observed magnitude and reddening for each relevant survey. To highlight the results of our model we show the color-color diagrams, with photometric uncertainties, for objects in each survey in Figure \ref{fig:CC}. The figure displays the exquisite photometry which will be acquired as part of each survey.

The results also highlight the separation between pure-Hydrogen and pure-Helium atmospheres. Since the Balmer lines are located in the $u$- and $g$-bands it is no surprise to see the populations well separated in the LSST and CASTOR samples.

Furthermore, Figure \ref{fig:CC} shows the density of points throughout the color-color diagram. Since white dwarfs cool as they age, redder colors generally indicate cooler, older, WDs. It can be seen that the densest location present in the NIR surveys occur at red temperatures, indicating a larger concentration of these cool objects relative to the hot WDs. In the optical/UV surveys, more uniform distributions can be seen, indicating a larger number of hot, young, WDs. These samples will be explored in detail in the following section.

\begin{figure*}[!t]
	\includegraphics[angle=0,width=0.49\textwidth]{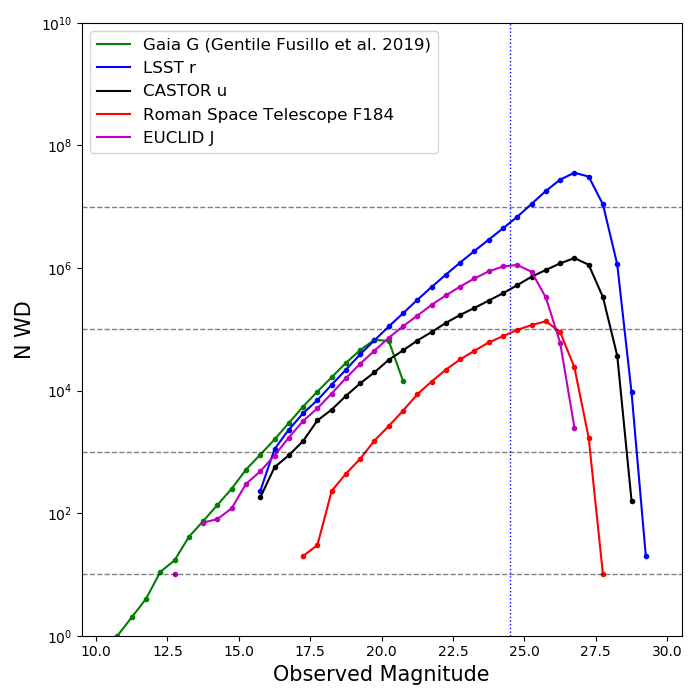}
	\includegraphics[angle=0,width=0.49\textwidth]{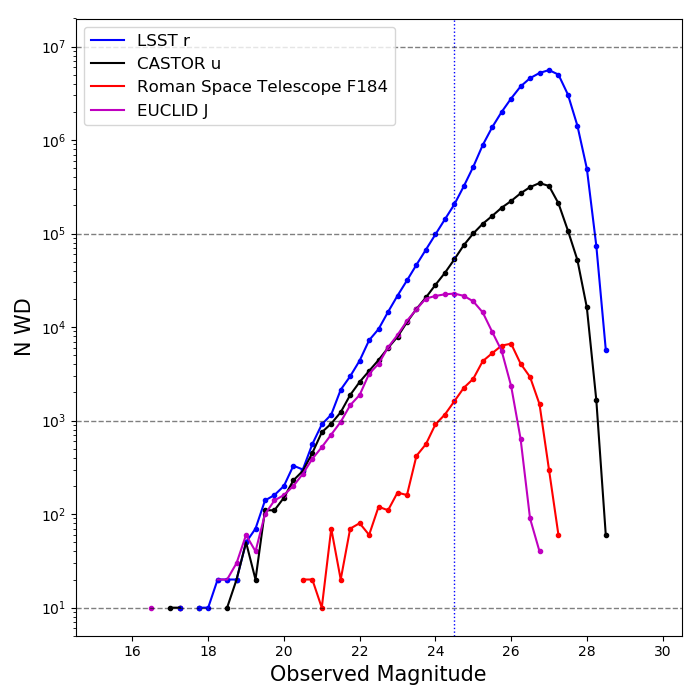}\\
	\includegraphics[angle=0,width=0.49\textwidth]{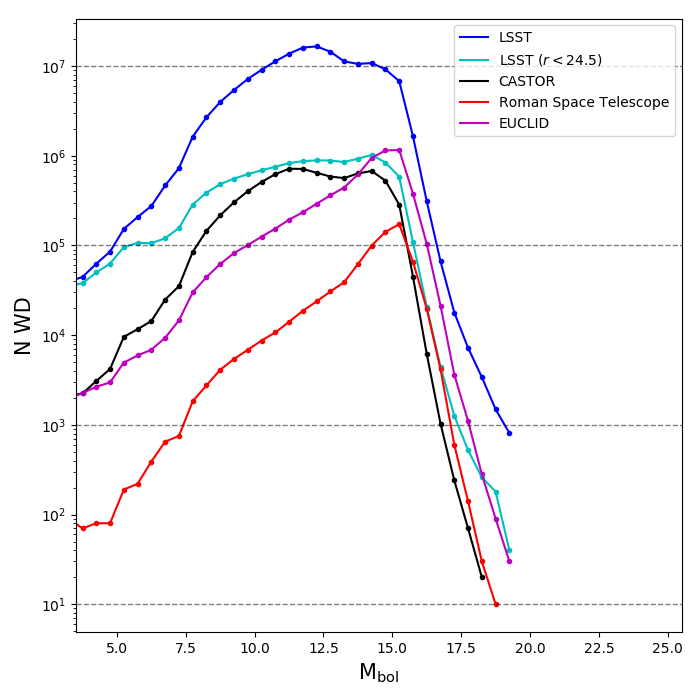}
	\includegraphics[angle=0,width=0.49\textwidth]{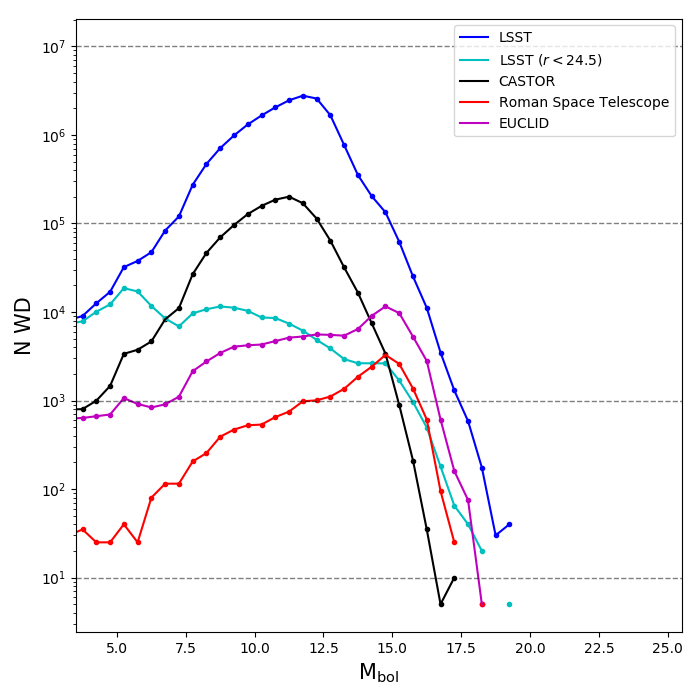}
	\caption{
	\textit{Top:} The observed magnitude distributions for all Galactic white dwarfs (left) found within each survey, with the halo component shown in the right column. These magnitudes include the effect of interstellar dust extinction as detailed in \S\ref{sec:model}. \textit{Bottom:} The bolometric magnitude distribution for each survey for all white dwarfs (left) and just the halo population (right). These distributions represent the temperature distribution of the simulated white dwarfs, showing that the hot white dwarfs will be observed by the UV surveys while the cool ones will be predominately observed by the IR surveys. \bigskip }
	\label{fig:mag_distribution}
\end{figure*}

\begin{figure*}[!t]
	\includegraphics[angle=0,width=0.49\textwidth]{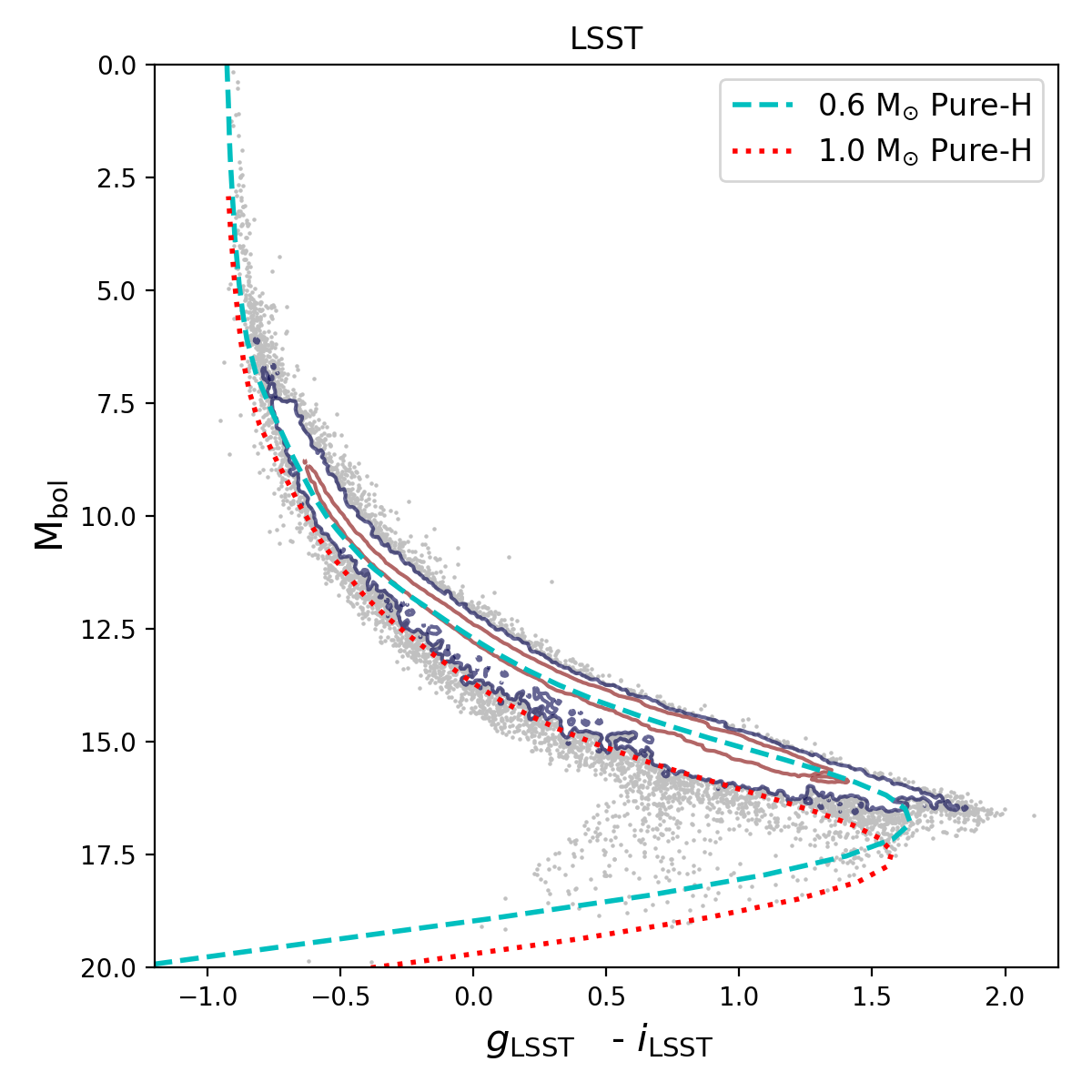}
	\includegraphics[angle=0,width=0.49\textwidth]{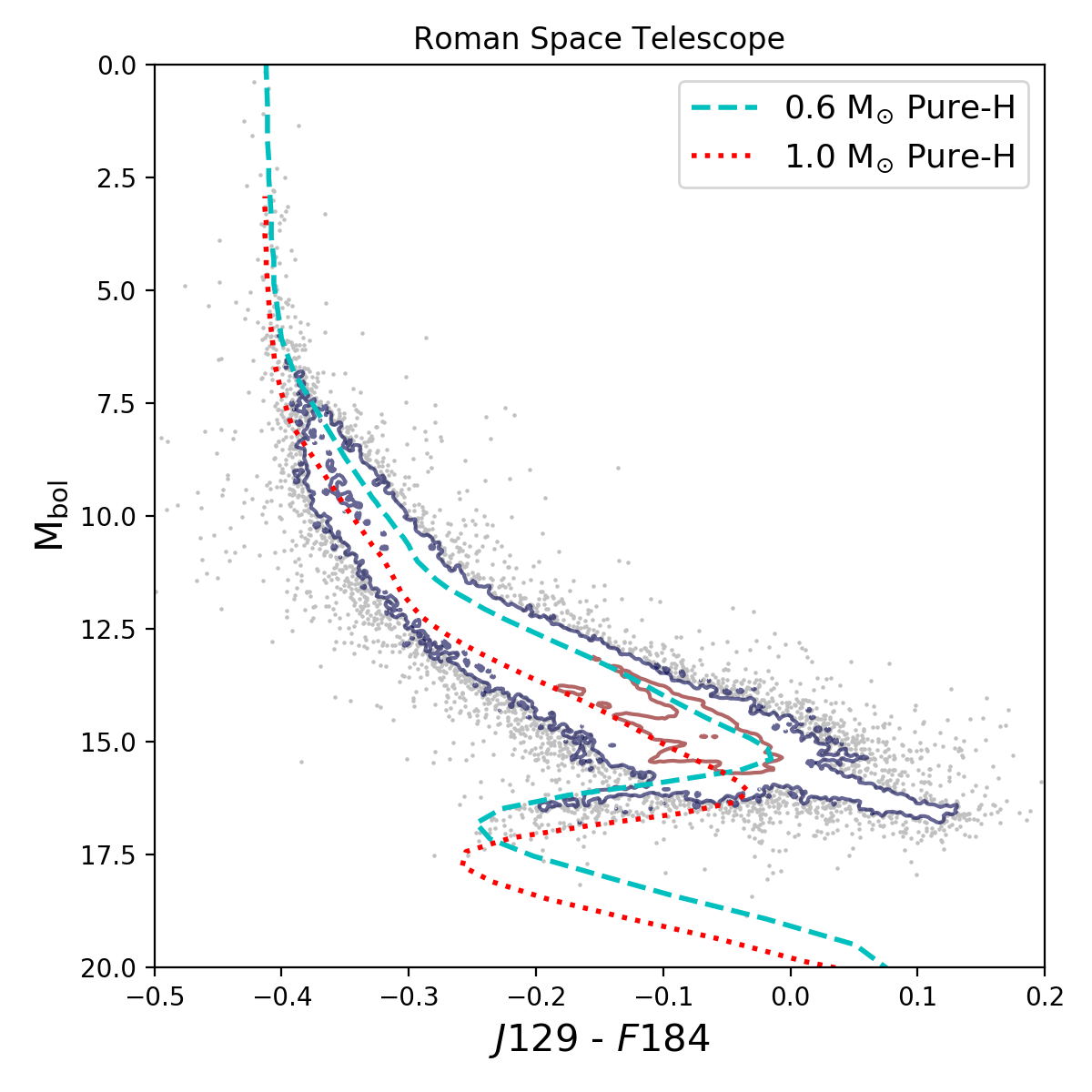}\\
	\includegraphics[angle=0,width=0.49\textwidth]{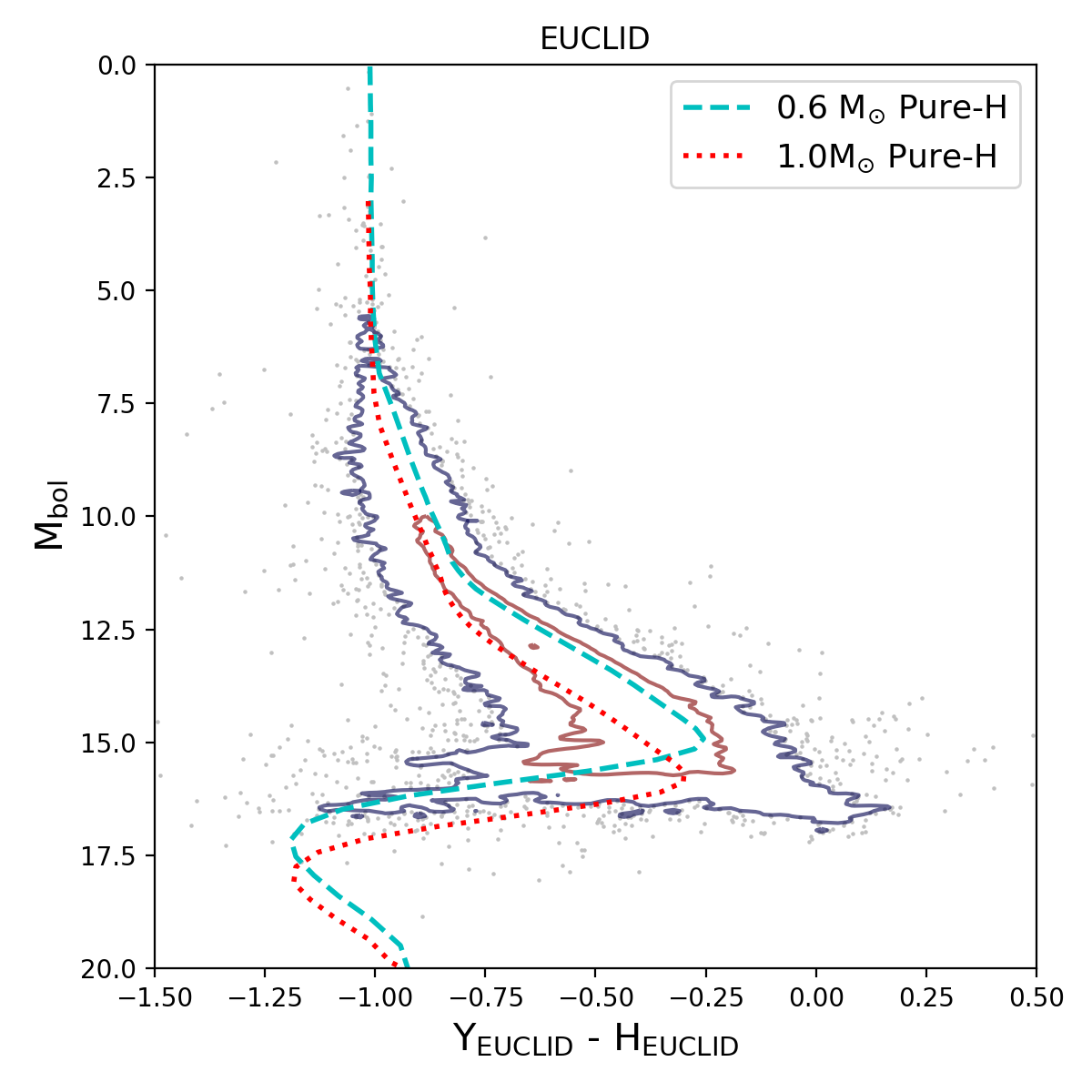}
	\includegraphics[angle=0,width=0.49\textwidth]{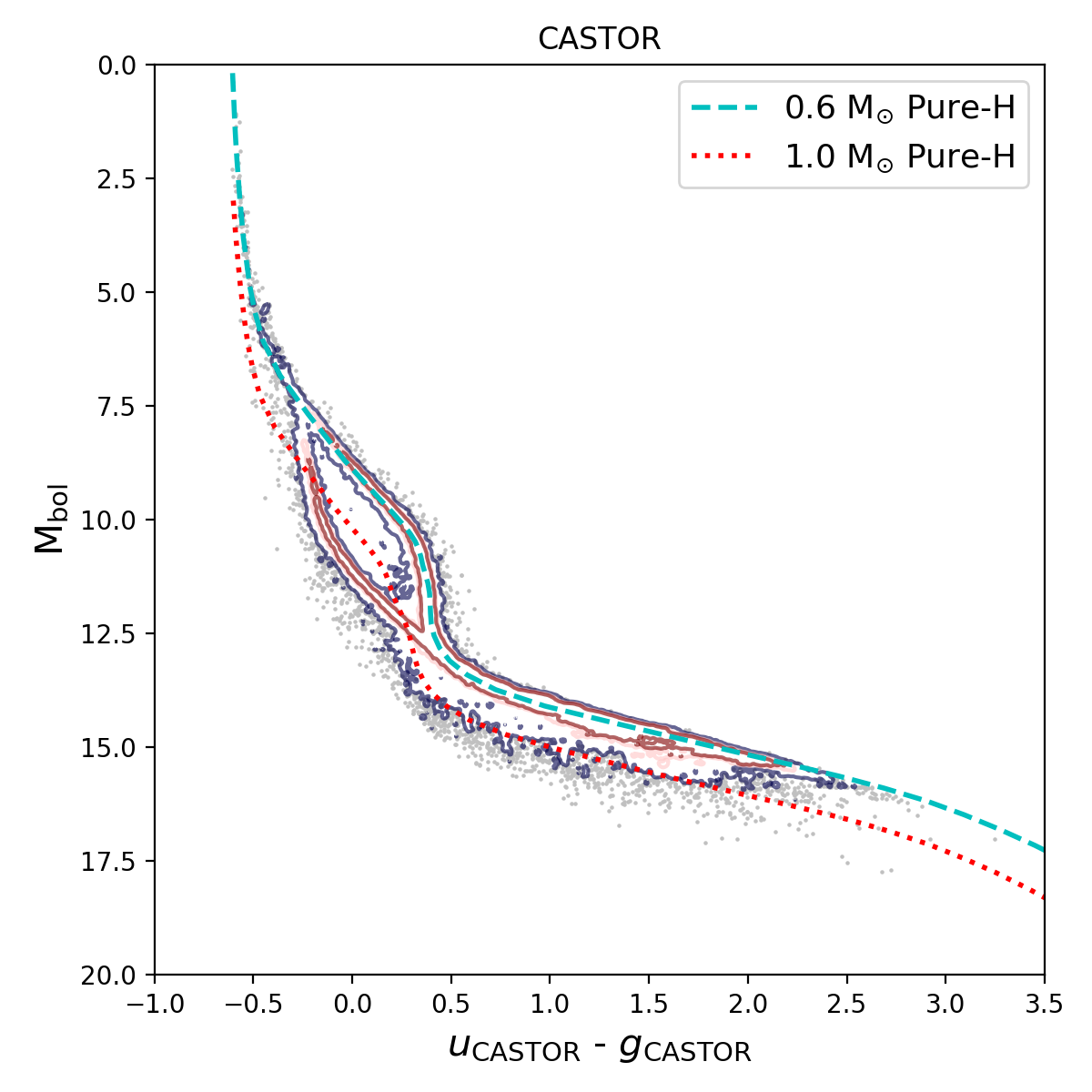}
	\caption{Color-magnitude diagrams for the four surveys with model 0.6\,M$_{\odot}$ pure-H (solid cyan) and 1.0\,M$_{\odot}$ pure-H  (dotted red) cooling curves over-plotted. The simulation includes the addition of photometric uncertainties.\bigskip }
	\label{fig:CMD}
\end{figure*}

\begin{figure*}[!t]
	\includegraphics[angle=0,width=0.99\textwidth]{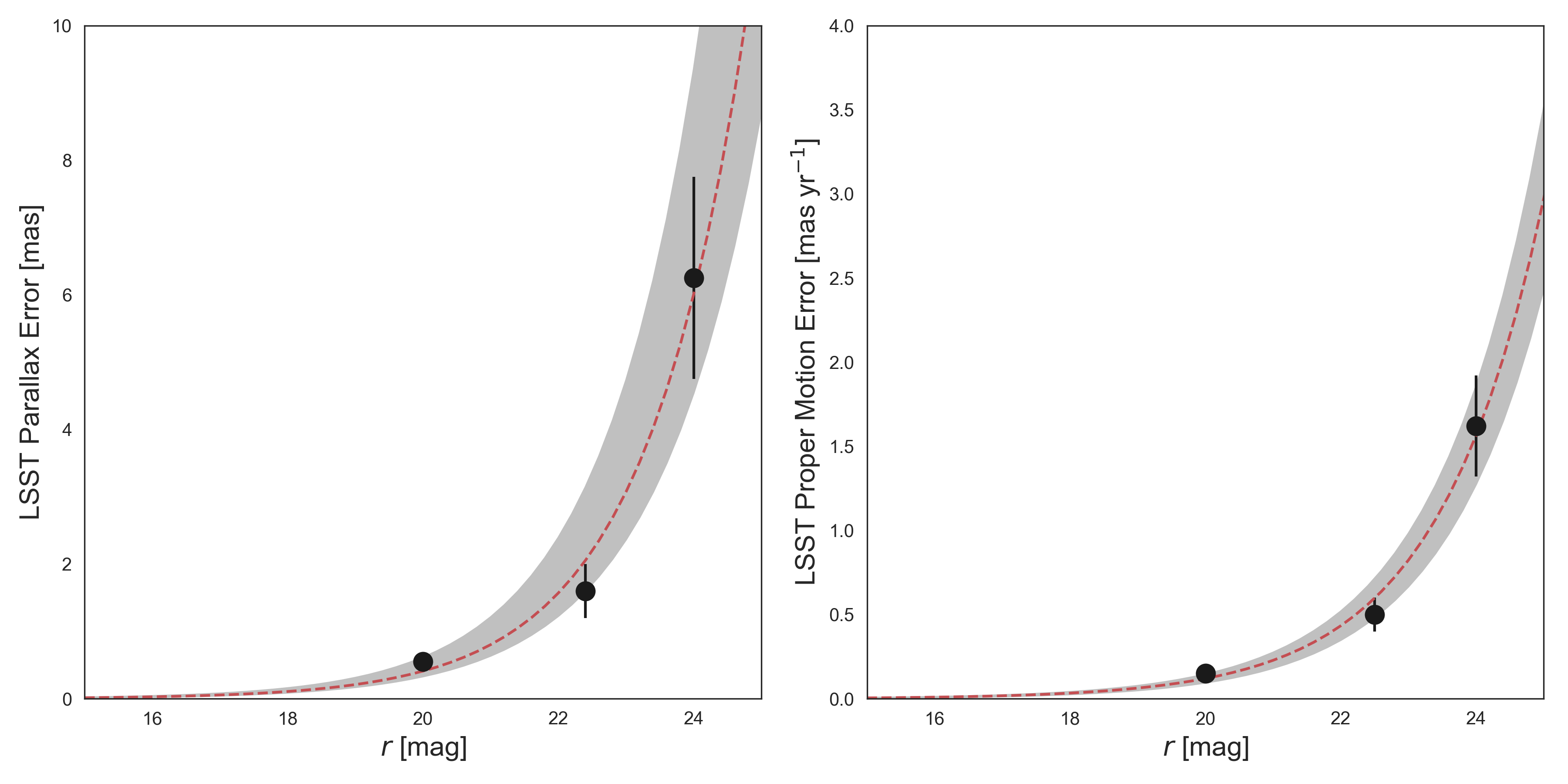}\\
	\includegraphics[angle=0,width=0.99\textwidth]{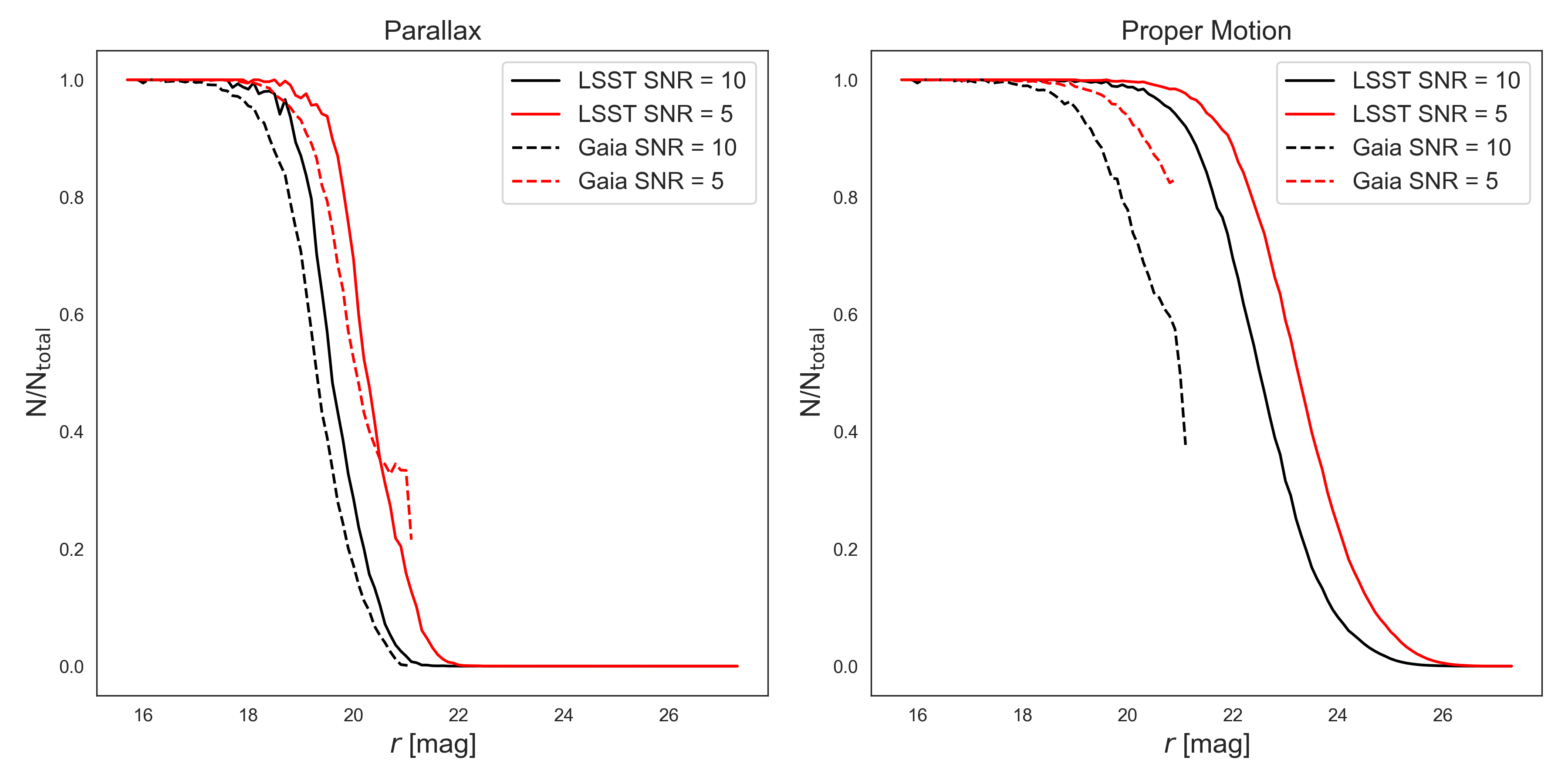}
	\caption{
	\textit{Top:} The astrometric precision from the LSST \textsc{OPSIM} is fit using an exponential function (grey region). The same is done for the proper motion precision as a function of observed $r$-band magnitude. \textit{Bottom:} The result of applying both of these prescriptions to the simulated dataset, showing the fraction of objects having greater than a 10$\sigma$ (black) and 5$\sigma$ (red) precision on the proper motion (bottom-right) or parallax (bottom-left). Also plotted are the results from the \textit{Gaia} WD sample presented in \cite{Gentile2019} (dashed lines). \bigskip }
	\label{fig:astrometry}
\end{figure*}

\section{Survey Results}
\label{sec:results}

Here we present the results of each simulation while highlighting the strengths of each survey, before discussing selection methods in the following section.

\subsection{Magnitude Distribution}

The magnitude distributions for each survey are presented for all WDs in the left-hand panel of Figure \ref{fig:mag_distribution}. Given the large area and extreme photometric depths that LSST will probe, it is no surprise to see this survey producing the greatest number of WDs. At the single epoch depth of $r\sim$24.5, the survey will observe on the order of 16 million WDs, and, at the full stacked depth of the 10-year survey, it will observe more than 150 million WDs.

Impressive samples will also result from CASTOR, the Roman Space Telescope, and Euclid. CASTOR will observe roughly 8 million WDs in the ultraviolet, allowing for the most complete sample of recently formed WDs, which will be important for the initial-to-final mass relation and fundamental physics like neutrino cooling \citep{Hansen2015}. The Roman Space Telescope will detect nearly a million WDs in its NIR bands, which will include the coolest and ancient WDs, as well as those with debris disks. Euclid will detect nearly 7 million WDs in its red-optical and NIR bands over a large area covering both hemispheres.

Also highlighted in the right-hand panel of Figure \ref{fig:mag_distribution} is the magnitude distributions for samples of halo WDs. LSST will detect the largest sample of halo objects, with approximately 50 million to full depth and nearly half a million to the single epoch depth of $r\sim$24.5. CASTOR's performance is also noteworthy, with nearly three million halo WDs observed within the survey. 

\subsection{Bolometric Magnitude Distributions}

In the bottom panels of Figure \ref{fig:mag_distribution} we also show the bolometric magnitudes of the WD samples. The bolometric magnitude is indicative of surface temperature, with fainter magnitudes representing cooler and older objects.

The plots highlight the relative strengths of the different surveys, particularly when it comes to halo objects. Surveys such as the Roman Space Telescope High Latitude Survey and Euclid, while observing fewer overall WDs, will observe more of the coolest and most ancient objects that lie beyond the turn-off in the WD luminosity function (see \S\ref{sec:discussion}). CASTOR and LSST will observe many of the newly formed WDs in the halo, and therefore the union of the four surveys will observe WDs covering a wide range in surface temperature.

\subsection{Color-Magnitude Diagrams}

In Figure \ref{fig:CMD} we show the resulting color-magnitude diagrams for each survey, while also highlighting the model cooling tracks for 0.6\,M$_{\odot}$ and 1.0\,M$_{\odot}$ pure-Hydrogen WDs. The results show that the NIR surveys, the Roman Space Telescope and Euclid, will observe a larger fraction of cool white dwarfs, while CASTOR will primarily see the hotter WDs. LSST, with its broad optical wavelength coverage and large area, will observe a significant number of both young and old WDs.

\section{Survey Selection Methods}

In this section, we describe the selection methods likely to be used within two footprints: the WFD area, and the Roman Space Telescope High-Latitude Survey. These two footprints will be the main focus of future WD studies, a few of which we highlight in the following section.

\subsection{The LSST WFD Survey Area}
\label{sec:selection}

The science goals explored in \S\ref{sec:discussion} will depend on our ability to select samples from the large WD populations presented in the previous section. Historically, WDs have been predominantly selected by either relying on photometry combined with parallaxes or proper motions. Of course, the efficiency of these methods depends on the precision of the derived values. To explore the samples expected to have adequate parallax and/or proper motion accuracy, we used results from the LSST survey simulator, \textsc{OPSIM} \citep{OPSIM, OPSIM1, OPSIM2}, and applied them to our model. 

The resulting distributions of parallax and proper motion precision can be seen in Figure \ref{fig:astrometry}, where the top panels show exponential fits to the results from \textsc{OPSIM} (black points). These functions are used to apply an uncertainty value to our model parallax and proper motions, and the bottom panels show the fraction of values, as a function of magnitude, above 5 and 10$\sigma$ precision. The {\it Gaia}-based WD catalog of \cite{Gentile2019} is plotted as a comparison, showing that the parallax precision for LSST will be similar to \textit{Gaia}, although the proper motions will extend well beyond the \textit{Gaia} magnitude limit. Below, we discuss the three ``selection regimes" for LSST WDs and present the resulting numbers in Table~\ref{table:number}.

\subsubsection{Regime 1: Selection on Parallax, Proper Motion, and Photometry}

The most accurate method for selecting WDs relies on using their parallaxes. Because they are intrinsically fainter than main-sequence stars at equal temperatures, WDs occupy a well-defined region in the color-magnitude diagram with minimal contamination from hot sub-dwarfs \citep[see][]{GaiaHRD}. This method was applied with great success in \textit{Gaia}, which measured parallaxes for more than a billion stars over the whole sky.

\cite{Gentile2019} used the \textit{Gaia} DR2 catalog to select WD candidates based on their position in the \textit{Gaia} color-magnitude diagram and imposed a variety of \textit{Gaia} quality flags to remove poor measurements. A comparison of this catalog and the \textsc{OPSIM} LSST simulation suggests that the LSST catalog will have similar precision to the \textit{Gaia} mission, albeit over somewhat less than half the total area.

A stricter cut performed specifically on the parallax signal-to-noise ratio (SNR) was used by \cite{Kilic2019} to measure accurate distances to a set of WDs having kinematics inconsistent with the Milky Way disk. The authors used a cut of 5$\sigma$ on the parallax as this reduced the uncertainty on their calculated radius, which is derived using multi-band photometry combined with the parallax. The bottom left-hand panel of Figure \ref{fig:astrometry} shows the 5$\sigma$ error curve for the parallax compared to the \textit{Gaia} dataset from \cite{Gentile2019}, showing roughly equal performance as a function of magnitude, with LSST sample extending a few tenths of a magnitude fainter. Thus, as with the \textit{Gaia} dataset, the majority of objects having $r\lesssim$ 20.0 mag will be able to make use of 5$\sigma$ parallax measurements.

\begin{table*}[!t]
\centering
\caption{Number of White Dwarfs in Each Survey}
\label{table:number}
\begin{tabular}{cc|c|c|c|c|c||c|c|c}

	\hline \hline
	Survey & Area & N$_{\textrm{WD}}$ & N$_{\textrm{WD}}$ & N$_{\textrm{WD}}$  &N$_{\textrm{WD}}$ & N$_{\textrm{WD}}$   N$_{\textrm{WD}}$ & N$_{\textrm{WD}}$ & N$_{\textrm{WD}}$  &N$_{\textrm{WD}}$    \\
	
	 & deg$^2$ &  & $\Delta\pi$ $>$ 10$\sigma$&  $\Delta\pi$ $>$ 5$\sigma$ & $\Delta\mu$ $>$ 10$\sigma$&  $\Delta\mu$ $>$ 5$\sigma$  &&$\Delta\pi$ $>$ 5$\sigma$ &$\Delta\mu$ $>$ 5$\sigma$\\
	 
	  &  &   &   &   &   & &HLS&HLS&HLS\\ 
	\hline
	
	LSST & 18,000  & 154,849,052  & 140,090  &  286,387 & 3,500,461   & 6,819,463   & 1,240,950 & 22,473 &360,229\\
	
	Euclid & 15,000  & 6,557,974  &  &   &   & &532,689&& \\
	
	Roman &  &   &   &   &  &   &&  & \\
	Space &  2,213  & 735,401  &   &   & 703,072 &  726,714   &735, 401& 703,072 &  726,714 \\
	Telescope &  &   &   &   &  &   &&  &   \\
	
	CASTOR & 7,700  & 7,796,149  &   &   &  &   &621,831&& \\
	
	\hline
	\bigskip
\end{tabular}

\end{table*}

\subsubsection{Regime 2: Selection on Proper Motion and Photometry}

Below \textit{Gaia}'s limiting magnitude of $r \sim$ 20, LSST's full impact will be achieved by providing proper motion measurements down to its single-epoch, 5$\sigma$ photometric limit of $r \sim$ 24.5.

The lower right-hand panel of Figure \ref{fig:astrometry} shows the fraction of objects, as a function of $r$-band magnitude, which will have greater than 5$\sigma$ (solid red) and 10$\sigma$ (solid black) measurements for their proper motion. We include the \textit{Gaia} DR2 sample of \cite{Gentile2019} for reference.

As with parallax measurements, the intrinsic faintness of WDs means that, at equal temperature or color, a WD will exhibit a larger proper motion than stars belonging to other stellar classes. This property has made it possible for clean samples of WDs within the SDSS footprint to be selected by \cite{Harris2006} and \cite{Munn2017} as well as in the SuperCOSMOS Sky Survey by \cite{RowellHambly2011}. Such methods rely on the reduced proper motion, $H$, of an object, which is defined as

\begin{equation}
\label{equation:RPM}
\begin{array}{lcl}
H & = & m + 5\log\mu + 5 \\
& = & M + 5\log v_{\mathrm{t}} - 3.379. \\
\end{array}
\end{equation}
Here $m$ is the magnitude in a given band and $\mu$ is the proper motion in arcseconds per year. $H$ can also be expressed in terms of the absolute magnitude, $M$, and the tangential velocity, $v_t$, measured in km~s$^{-1}$.

The studies of \cite{RowellHambly2011} and \cite{Munn2017} used proper motion cuts at 5$\sigma$ and 3.5$\sigma$, respectively, to select their WD samples. As Figure \ref{fig:astrometry} shows, the approximate magnitude limit for which LSST will recover 50\% of objects with proper motion measurements greater than 5$\sigma$ is $r \sim$ 23.2. This limit is 2-4 magnitudes deeper than the aforementioned studies. Selecting all objects with 5$\sigma$ or better accuracy on their proper motion measurement results in nearly 7 million WDs, two orders of magnitudes larger than previous studies.

\subsubsection{Regime 3: Selection on Photometry}

The majority of WD detected by LSST will be beyond the limits needed for accurate parallax and proper motion measurements. These objects, more than 140 million of them, will nevertheless have accurate six-band photometry down to the full depth of the survey. The primary obstacle with identifying WDs at these faint magnitudes is that the WD cooling sequence will overlap with many other stellar classes within a color-color diagram, particularly at cooler temperatures. 

Photometric selection of WDs, therefore, focuses on hot WDs within a color-color diagram. A selection of this manner was performed in \cite{Fantin2017} using data from the Next Generation Virgo Cluster Survey finding a contamination rate of 14\% for WDs with temperatures above 12,500\,K. This contamination rate, however, was measured using SDSS spectroscopy which is incomplete and typically contains objects brighter than $r$=21.0 mag. A much smaller contamination rate (nearly zero) was found by \cite{Bianchi2011} using GALEX UV photometry, which again was determined to the limit of the SDSS spectroscopic capabilities.

While we cannot extrapolate to the depth of the LSST, the rate of contamination will certainly be much higher when imaging nearly 7 magnitudes fainter. This regime will need to rely on the full spectral energy distribution (SED) of the objects to determine the probability of an object being a WD. Several large spectroscopic surveys, such as WEAVE, DESI and 4MOST, will aid in the creation of large, diverse training sets for such endeavors.

\subsubsection{Star-Galaxy Separation}

One important caveat to these selection methods is the ability to separate Milky Way stars from background galaxies which become more distant and compact at fainter magnitudes. For example, a recent study by \cite{Slater2020} showed that the number of background galaxies is equal to the number of stars at $r\sim$20.5 and increases to  nearly 13:1 at $r$=24.0 and 25:1 at $r$=25.0. This means that accurate star galaxy separation will be important for Regime 2, but critically so for Regime 3. 

Since background galaxies should have zero proper motion, the selection of WDs, which exhibit larger proper motions than most stellar populations, should be possible up to the faint end of Regime 2 given the 5$\sigma$ proper motion accuracy. As \cite{Slater2020} showed, LSST should be able to classify 80\% of stars as being stars down to a limit of $r$-band magnitude of 23 - 23.5 based solely on shape parameters derived from images. Combining this information with  proper motion measurements should allow WDs to be well distinguished from the background galaxies when accurate proper motions are available.

In Regime 3, however, shape measurements combined with the full 6-band photometry will need to be used to separate WDs from background galaxies, and the efficacy of such an approach is beyond the scope of this paper \citep[for work on this topic, see e.g,][]{Fadely2012, Soumagnac2015, Kim2017, Bai2019, Slater2020}.

\begin{figure*}[!t]
	\includegraphics[angle=0,width=0.49\textwidth]{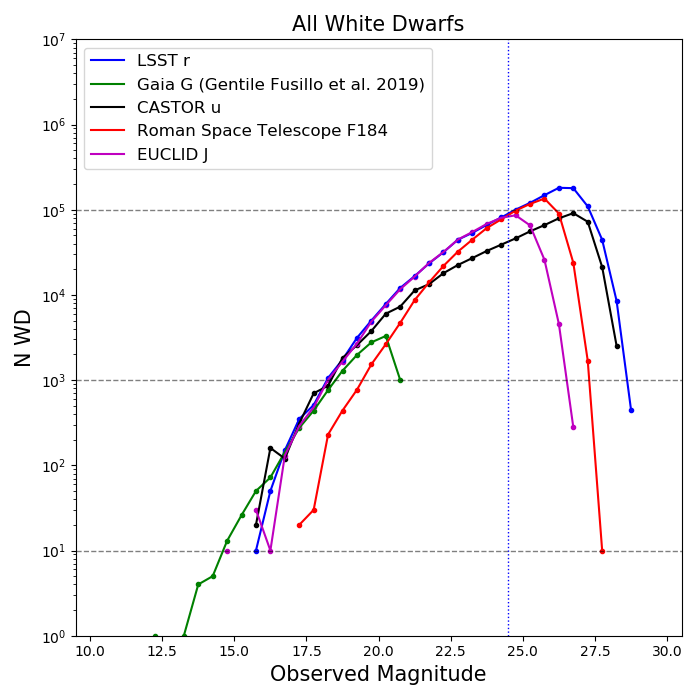}
	\includegraphics[angle=0,width=0.49\textwidth]{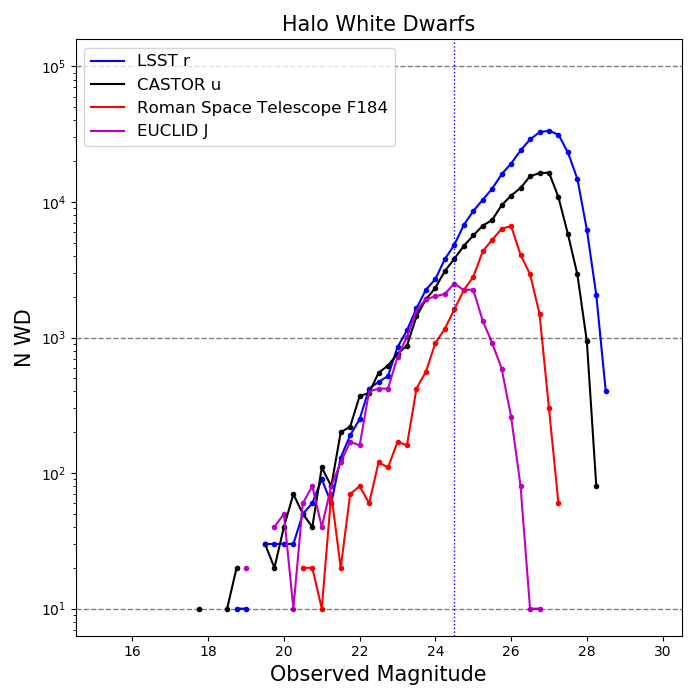}\\
	\includegraphics[angle=0,width=0.49\textwidth]{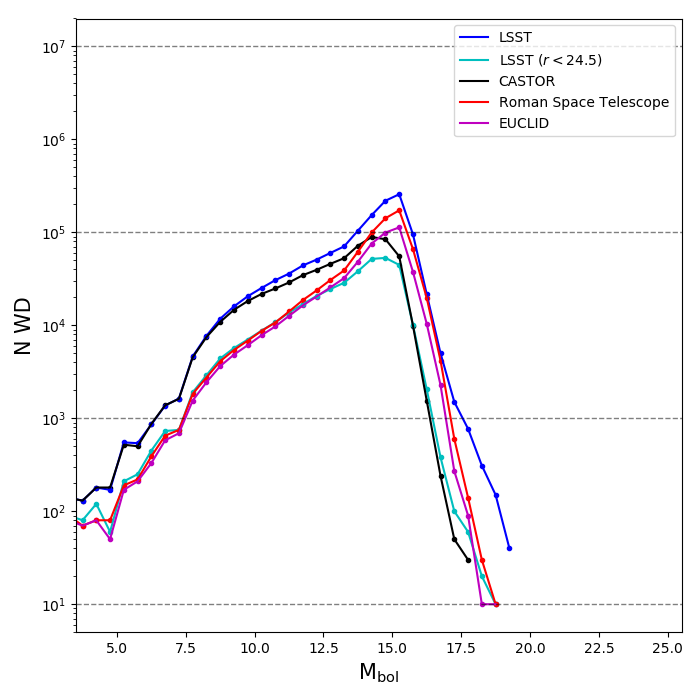}
	\includegraphics[angle=0,width=0.49\textwidth]{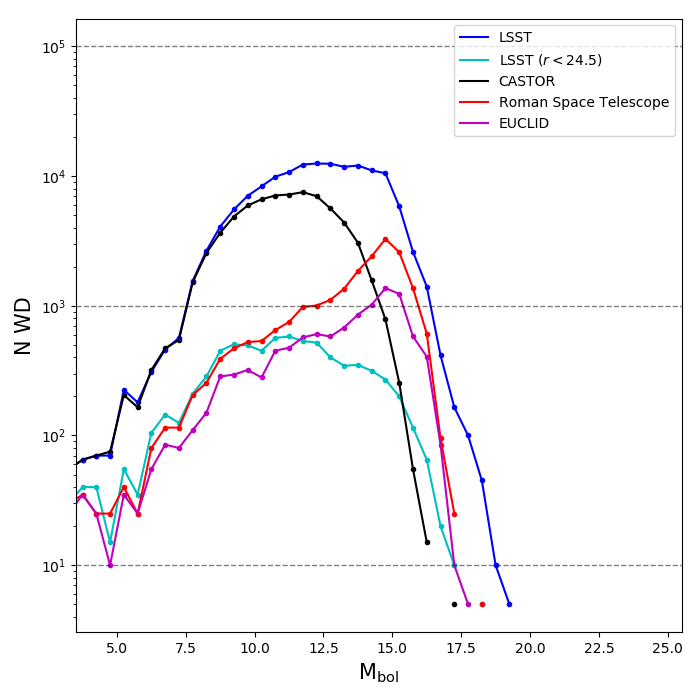}
	\caption{
	\textit{Top:} The observed magnitude distributions for all white dwarfs (left) found within the Roman Space Telescope HLS, with the halo highlighted in the right column. These magnitudes include the effect of interstellar dust extinction as detailed in Section \ref{sec:model}. \textit{Bottom:} The bolometric magnitude distribution for each survey within the Roman Space Telescope HLS for all white dwarfs (left) and just the halo population (right). \bigskip }
	\label{fig:mag_distribution_HLS}
\end{figure*}

\subsection{The Roman Space Telescope High Latitude Survey}
\label{sec:HLS}

While LSST will soon uncover a huge number of WDs, this sample will be made even more powerful by taking advantage of its overlap with other surveys. Space-based facilities like Euclid, the Roman Space Telescope, and CASTOR will dramatically improve star-galaxy separation over large portions of the WFD survey area since these facilities will have nearly 5$\times$ better image quality. Moreover, these space-based surveys will provide photometric data at ultraviolet and NIR wavelengths, allowing for more complete characterization of WD SEDs and the improved selection of double-degenerate and WD/main-sequence binaries. Meanwhile, multi-epoch observations will make it possible to derive relative proper motions between the surveys.

To highlight the research opportunities that would be made possible by multi-wavelength and multi-epoch observations within a large survey region common to all facilities, we consider the Roman Space Telescope High-Latitude Survey (HLS). This survey spans an area of 2,200 deg$^2$ in the Southern Sky, located far from the Galactic disk (b $< -$40, see Figure \ref{fig:maglimits}). While the goal of the HLS is to study dark energy by measuring the redshifts of more than 20 million galaxies and the shapes of more than 500 million galaxies, the deep multi-band infrared space-based imaging will provide a wealth of Galactic science cases as well \citep{WFIRST}. 

\begin{figure*}[!t]
	\includegraphics[angle=0,width=0.49\textwidth]{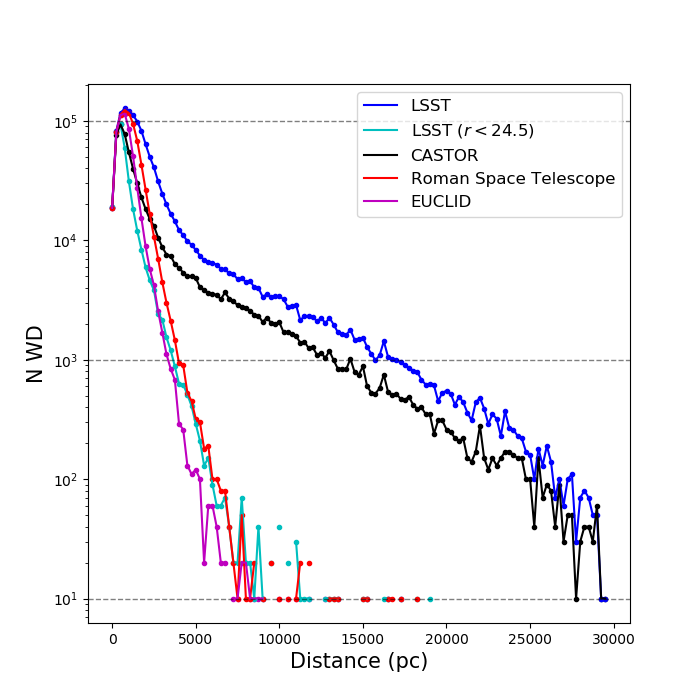}
	\includegraphics[angle=0,width=0.49\textwidth]{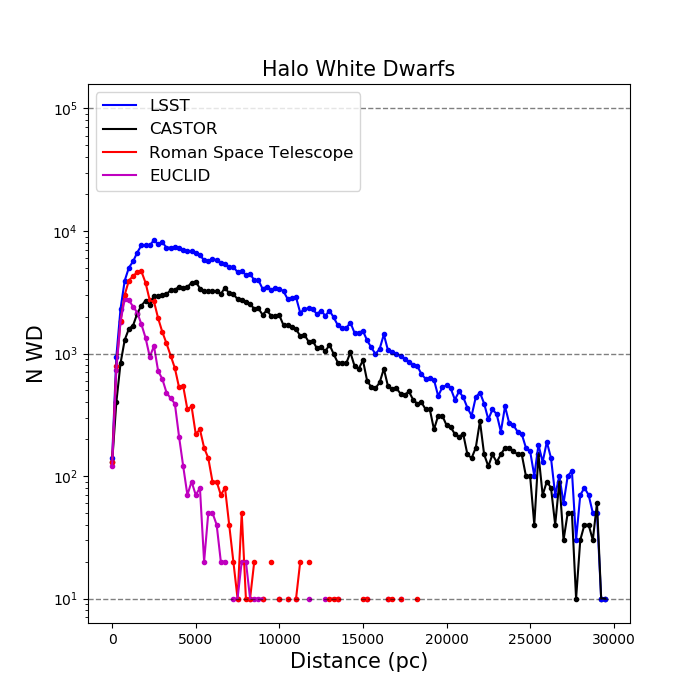}
	\caption{
	\textit{Left:} Distance distribution of all white dwarfs within the High Latitude Survey (HLS) field seen by the various surveys. The UV surveys will observe hot white dwarfs over a much larger volume owing to their higher intrinsic brightness.  \textit{Right:} Halo white dwarf distance distribution in the HLS field showing a similar result. \bigskip }
	\label{fig:distance}
\end{figure*}

The Galactic science cases will be bolstered by the addition of kinematic information to the depths of the HLS. According to current plans, the Roman Space Telescope HLS will be executed over 5 years which should allow for the measurement of proper motions. This topic has been explored by \cite{Sanderson2017} who estimated that the Roman Space Telescope will be able to produce relative proper motions with a precision 25\,$\mu$as/yr to the depth of the survey ($J_{\rm AB} \simeq 26.9$~mag). This value is roughly four times better than proper motions achieved over a similar baseline \textit{HST} \citep[100\,$\mu$as/yr, see, e.g,][]{Deason2013}. However, if only a precision similar to \textit{HST} is achieved, then it will still revolutionize the selection of WDs. The ability to combine positions across surveys, the improved precision that may result from an extended \textit{Gaia} mission, and optimized scheduling, could further improve the actual astrometric precision while allowing for an accurate transformation from relative to absolute proper motions \citep{Sanderson2017}.

One advantage that cool WDs will have is that their intrinsic faintness means that they will be located much closer to the Sun than other objects observed at the limit of the survey. This means that these WDs will experience a much larger proper motion, and as a result, the SNR of any measured proper motion measured wholly within the survey will be higher. For example, a 3,500\,K halo white dwarf at a distance of 750\,pc present in our model will still experience a proper motion of 42\,mas/yr, resulting in a total motion of 210\,mas throughout the 5\,year nominal survey. With a precision of 1.1\,mas in position, this object will move nearly 200x this value. Similar objects (M$_{bol}\sim$15-17) experience a typical proper motion of 12\,mas/yr in the thin disk, 22\,mas/yr in the thick disk, and 50\,mas/yr in the halo, moving a combined 60-250\,mas throughout the survey --- well above the positional precision of a single epoch measurement.

Proper motions also present a challenge when cross-matching two surveys taken at different epochs as the WDs can move significantly in the time between the surveys. Thus, these motions will need to be taken into account when performing the cross-matching with methods such as the one presented in \cite{Gentile2017}. This method uses the proper motion to calculate the projected position in the survey which does not include astrometric measurements to reduce mismatched objects, and has recently been applied to the SDSS-\textit{Gaia} catalog presented in \cite{Gentile2019}. Since the HLS will likely have proper motions to the depth of the survey, the ability to cross-match this sample with other catalogs will need to rely on these measurements.

If we apply the astrometric precision from \cite{Sanderson2017} to our sample of HLS WDs, we find that 99\% of the WDs in the HLS would have a proper motion detected at a 5$\sigma$ level or better. This would allow a selection comparable to that of Regime 2 for the LSST to the full depth of the HLS ($J_{\rm AB}\sim$26.9), allowing many of the oldest and coolest WDs to be detected with ease. We first present the results of the HLS simulation before exploring the implications of these results in the following section.

Figure \ref{fig:mag_distribution_HLS} shows the results of our WD simulations for the HLS, with the halo component highlighted in the right-hand column. The numbers within each sample are also included in Table \ref{table:number}. With each sample distributed over the same survey area, the difference in survey depths becomes apparent in the top panel of Figure \ref{fig:mag_distribution_HLS}.

The bolometric magnitude distributions reflect the passbands used in each survey since WDs have SEDs similar to blackbodies (see Figure \ref{fig:maglimits}). The NIR surveys, the Roman Space Telescope and Euclid, will uncover a large number of cool WDs, whereas the optical/UV surveys, CASTOR and LSST, will uncover the hotter, more recently formed WDs. The Roman Space Telescope alone will uncover a few thousand halo WDs beyond the turn-off in the WD luminosity function, making it the go-to survey within the HLS footprint when looking to determine the age and star formation of the Galactic halo. The ability to cross-match these surveys over such a large area means that WDs of all temperatures will be observed, providing a more complete census of the true Galactic population.

In Figure \ref{fig:distance}, we show the distance distribution of WDs detected within the HLS field. Given that the Roman Space Telescope and Euclid surveys mainly target the intrinsically faint, cool WDs, the volume of space which they survey is relatively small compared to CASTOR and LSST. CASTOR, in particular, will detect young WDs up to distances of 30 kpc, although most will lie within a few kpc of the Sun, still far beyond current capabilities.

Our results show that within the HLS footprint, the selection of WDs will rely heavily on the proper motions derived by the Roman Space Telescope, as the combination of precision and depth will surpass that of the LSST. Furthermore, the superior image quality will greatly aid in star-galaxy separation at the limits of the survey, allowing for a clean sample selection. With the added depth it is also possible to select cool WDs at a greater distance than within the WFD survey, although, given the intrinsic faintness of these objects, even the distances probed by the Roman Space Telescope will be mainly within 1\,kpc of the Sun.

\section{White Dwarf Science Cases}
\label{sec:discussion}

With an understanding of the different catalogs and their limitations, this section explores some key WD science cases. We emphasize that these are by no means comprehensive. Various science cases make use of different samples as defined in \S\ref{sec:selection} and \ref{sec:HLS} with the absolute numbers within both the WFD and the HLS footprints presented in Table \ref{table:number}.

\subsection{The White Dwarf Luminosity Function}
\label{subsec:LF}

\begin{figure*}[!t]
	\includegraphics[angle=0,width=0.49\textwidth]{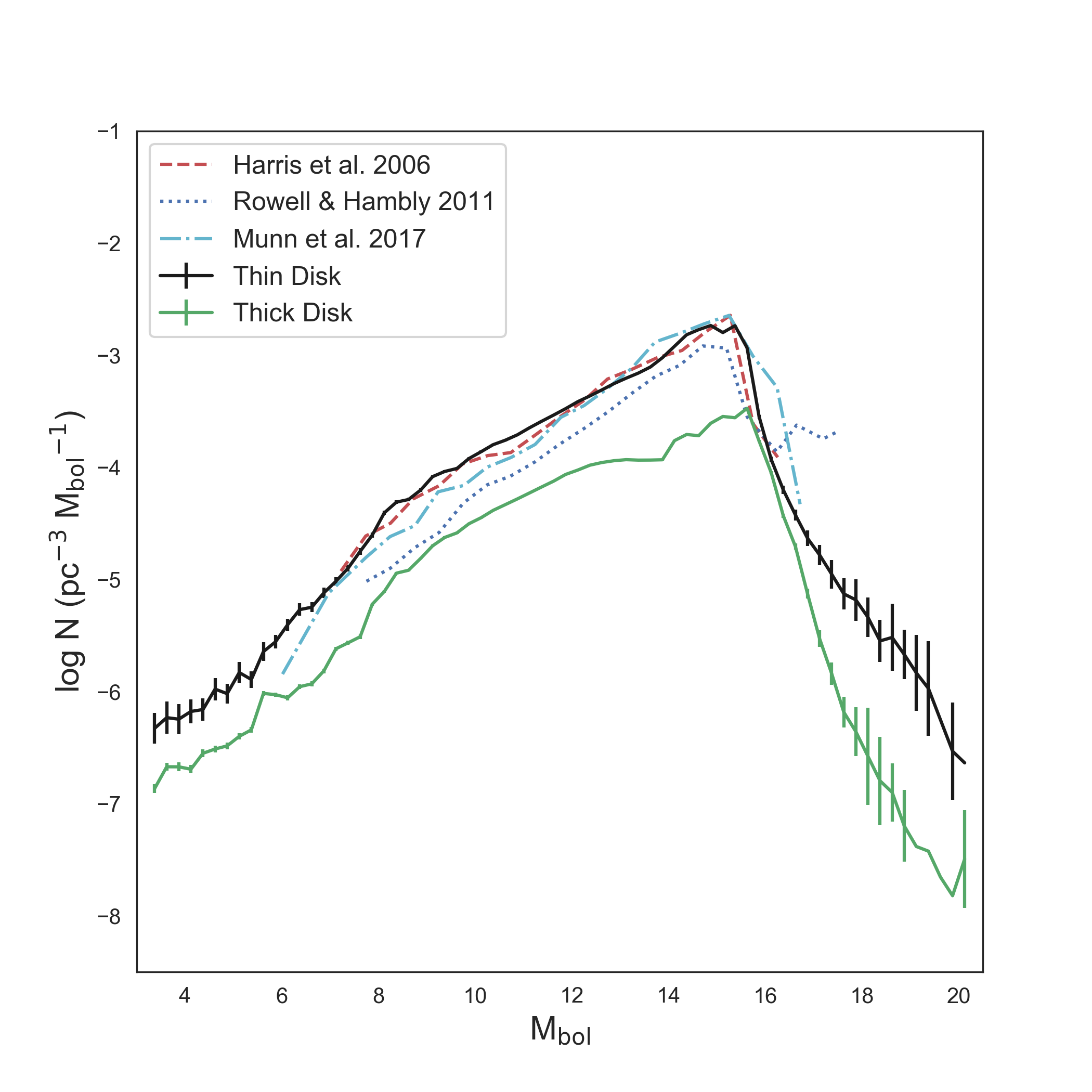}
	\includegraphics[angle=0,width=0.49\textwidth]{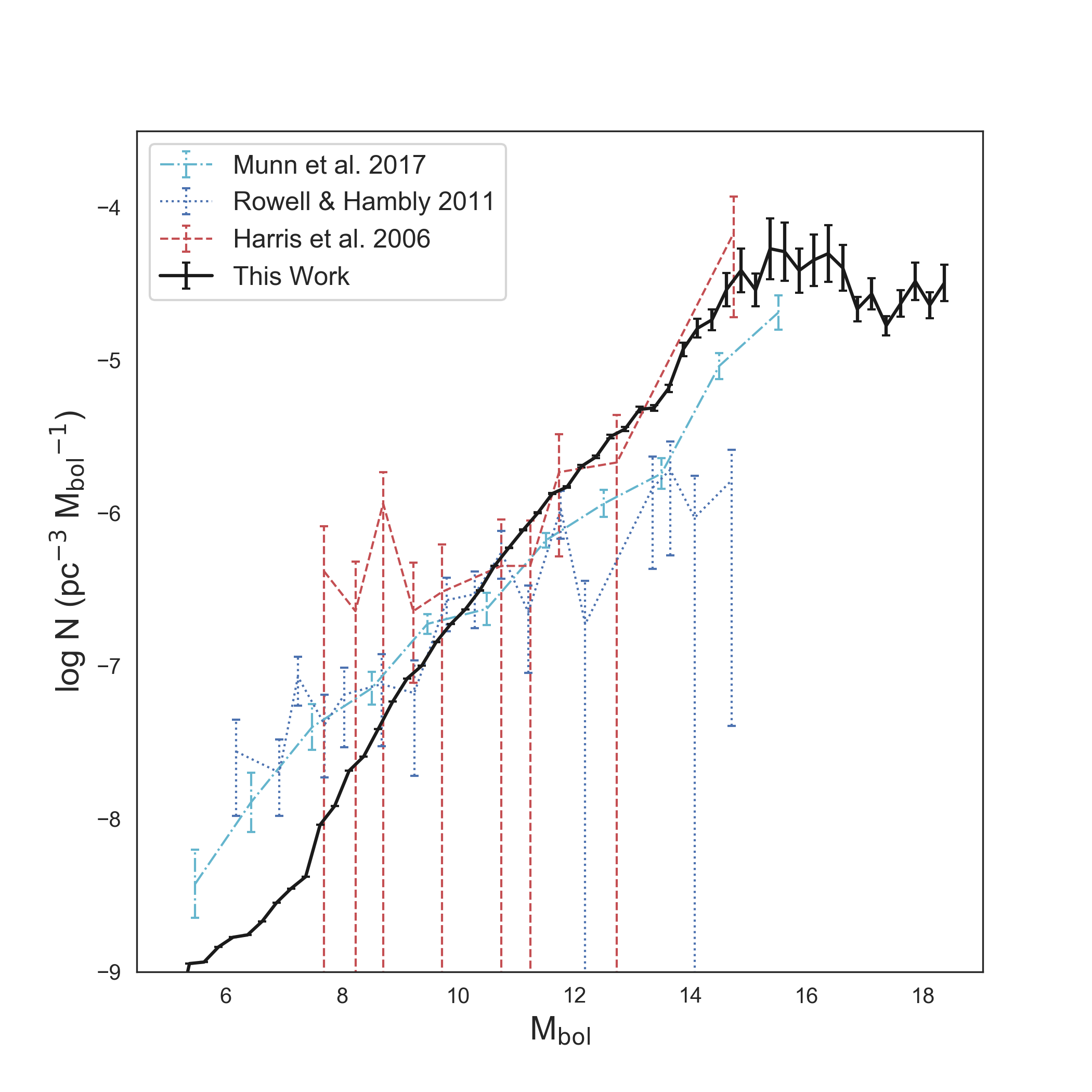}
	\caption{ \textit{Left:} The luminosity function for thin and thick disk white dwarfs found within the LSST 5$\sigma$ proper motion sample. Also plotted are results from \cite{Harris2006}, \cite{RowellHambly2011}, and \cite{Munn2017}. \textit{Right:} The halo white dwarf luminosity function with the same cut in proper motion precision. \bigskip }
	\label{fig:LF}
\end{figure*}

The WD luminosity function (WDLF) has been used in the past to determine ages and star formation histories for a variety of stellar populations, from the solar neighborhood to more distant star clusters\citep[see, e.g,][]{Richer2006,Bedin2008,Bedin2009,Rowell2013,Torres2015, Salaris2018}. This is a consequence of a decrease in the cooling rate of WDs at low temperatures combined with the finite age of the Universe. The result is that WDs become more numerous at low temperatures before dropping off below this regime as objects will not have had enough time to cool further. This manifests itself as a turn-off in the WDLF. An early example of this phenomenon is the study by \cite{Winget1987}, who used the turn-off in the local population of WDs to conclude that the age of the Galactic disk is 9.3\,$\pm$\, 2.0\,Gyr.

Following the large increase in the number of known WDs from surveys such as the SDSS, the WDLF has been determined with greater precision as the fainter magnitude limits increased the number of WDs found beyond the observed turn-off. For example, \cite{Harris2006} found four stars beyond the turn-off at M$_{\mathrm{bol}}$ = 15.25, and \cite{Munn2017} found on the order of a hundred. This resulted in \cite{Kilic2017} providing the most accurate age determination of the Milky Way's components to date using white dwarfs. Their results imply that the thin disk formed 7.4-8.2\,Gyr ago, the thick disk at 9.5-9.9\,Gyr, and the halo at 12.5$_{-3.4}^{+1.4}$, although a turn-off in the halo luminosity function has yet to be observed. The addition of proper motion data from \textit{Gaia} DR2, in combination with photometric data from CFIS and PS1, was used by \cite{Fantin2019} to determine the onset of star formation in the disk to be (11.3 $\pm$ 0.5)\,Gyr before reaching a maximum star-formation rate at (9.8 $\pm$ 0.3)\,Gyr, followed by a slight decline towards the present day. Given that the turn-off is the fundamental part of age determinations, finding objects beyond this turn-off is paramount and will require deeper surveys with accurate selection methods such as the WFD or HLS samples.

In Figure \ref{fig:LF} we show the predicted WDLF for the LSST WFD survey using objects with 5$\sigma$ proper motion measurements and compare it to results in the literature. Our WDLF was calculated using the 1/V$_{\mathrm{max}}$ method, which has been used for previous studies of observational WDLFs. This method, introduced by \cite{Schmidt1968}, aims to calculate the number density by summing the maximum volume each observed object could occupy and still be observed based on the survey parameters. Here, $V_{\mathrm{max}}$ is defined as 
\begin{equation}
    V_{\mathrm{max}}  =  \beta \int_{r_{\mathrm{min}}}^{r_{\mathrm{max}}} \frac{\rho}{\rho_{\odot}}R^{2}\mathrm{d}R, 
\end{equation}
where $\beta$ is the solid angle subtended by the survey, $\frac{\rho}{\rho_{\odot}}$ is the density distribution of the component and 
\begin{equation}
\centering
r_{min}  =  10^{0.2(m_{\mathrm{min}} - M)} 
\end{equation}
\begin{equation}
r_{max}  =  10^{0.2(m_{\mathrm{max}} - M)} 
\end{equation}
are the minimum and maximum distances for which the object would be observed given the bright and faint magnitude limits of the survey.

We use the density distributions as presented in \S\ref{sec:model} while the solid angle is calculated using the survey parameters. The number density is then the sum of the maximum volumes for each object within each bolometric magnitude bin,
\begin{equation}
\phi  =  \sum_{i=1}^{N}\frac{1}{V_{\mathrm{max}, i}} 
\end{equation}
while the uncertainty follows Poissonian statistics for each bin,
\begin{equation}
\sigma_{\phi}^2  =  \sum_{i=1}^{N}\frac{1}{V_{\mathrm{max}, i}^2}. 
\end{equation}

Our results show a disk luminosity function that increases from M$_{\mathrm{bol}}$ = +4 to +15.25~mag before dropping off sharply. This drop-off is a consequence of the finite age of the universe combined with the mass distribution of WDs. Since high-mass C/O WDs cool faster than their low-mass counterparts, we see a significant number of high-mass WDs beyond this turn-off, even in the younger thin disk population. These objects, with masses around 0.8--1.0\,M$_{\odot}$, are the remnants of intermediate-mass stars formed between 8 and 12\,Gyr. For the LSST, we find there will be nearly 100,000 of these objects in the 5$\sigma$ proper motion sample.

Our halo luminosity function is presented in the right-hand panel of Figure \ref{fig:LF}. The results show that the increased depth of the LSST will detect the turn-off in the halo luminosity function for the first time, with more than 40,000 objects below M$_{\mathrm{bol}}$ = 15.5.

\subsection{The Initial-to-Final Mass Relation}

\begin{figure*}[!t]
	\includegraphics[angle=0,width=0.33\textwidth]{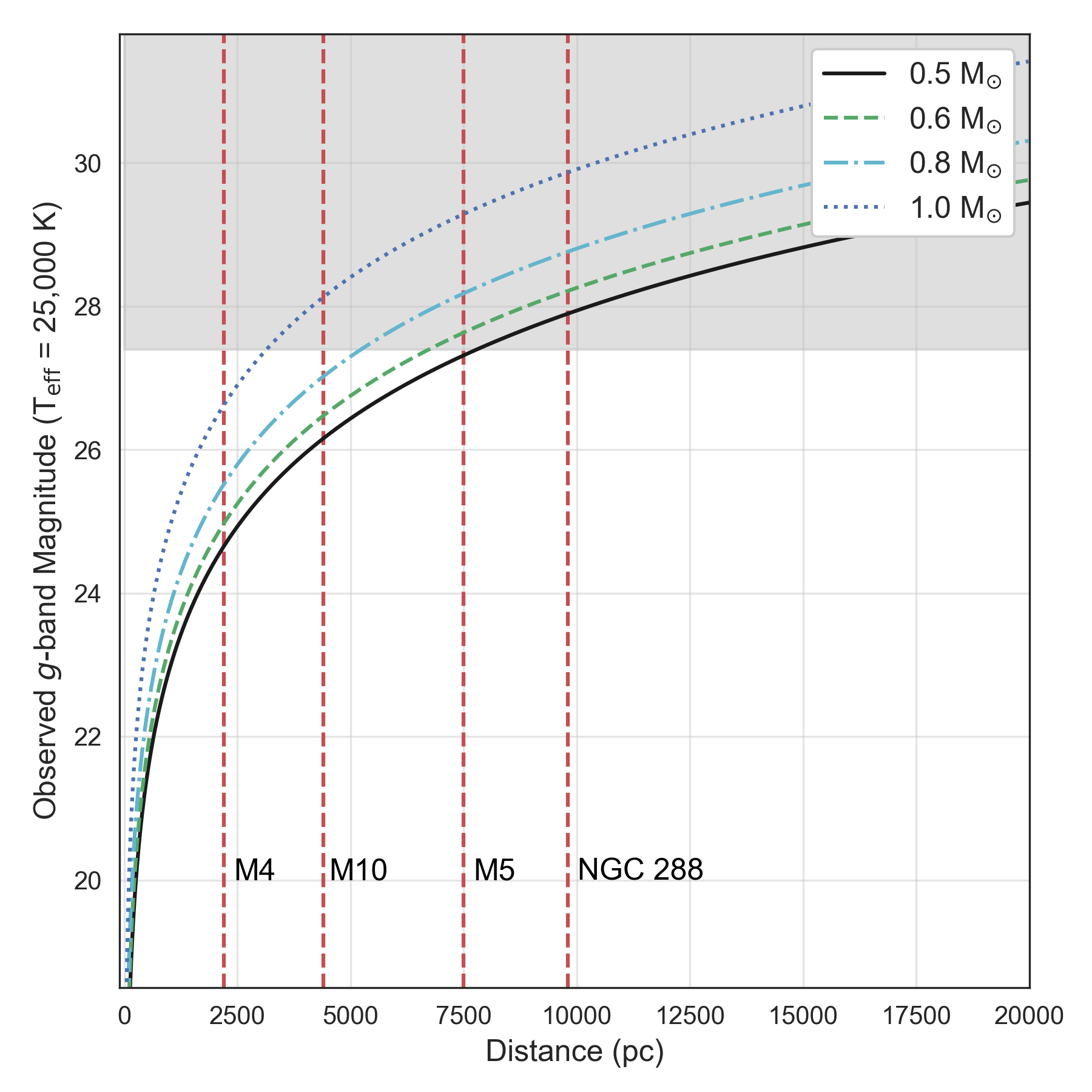}
	\includegraphics[angle=0,width=0.33\textwidth]{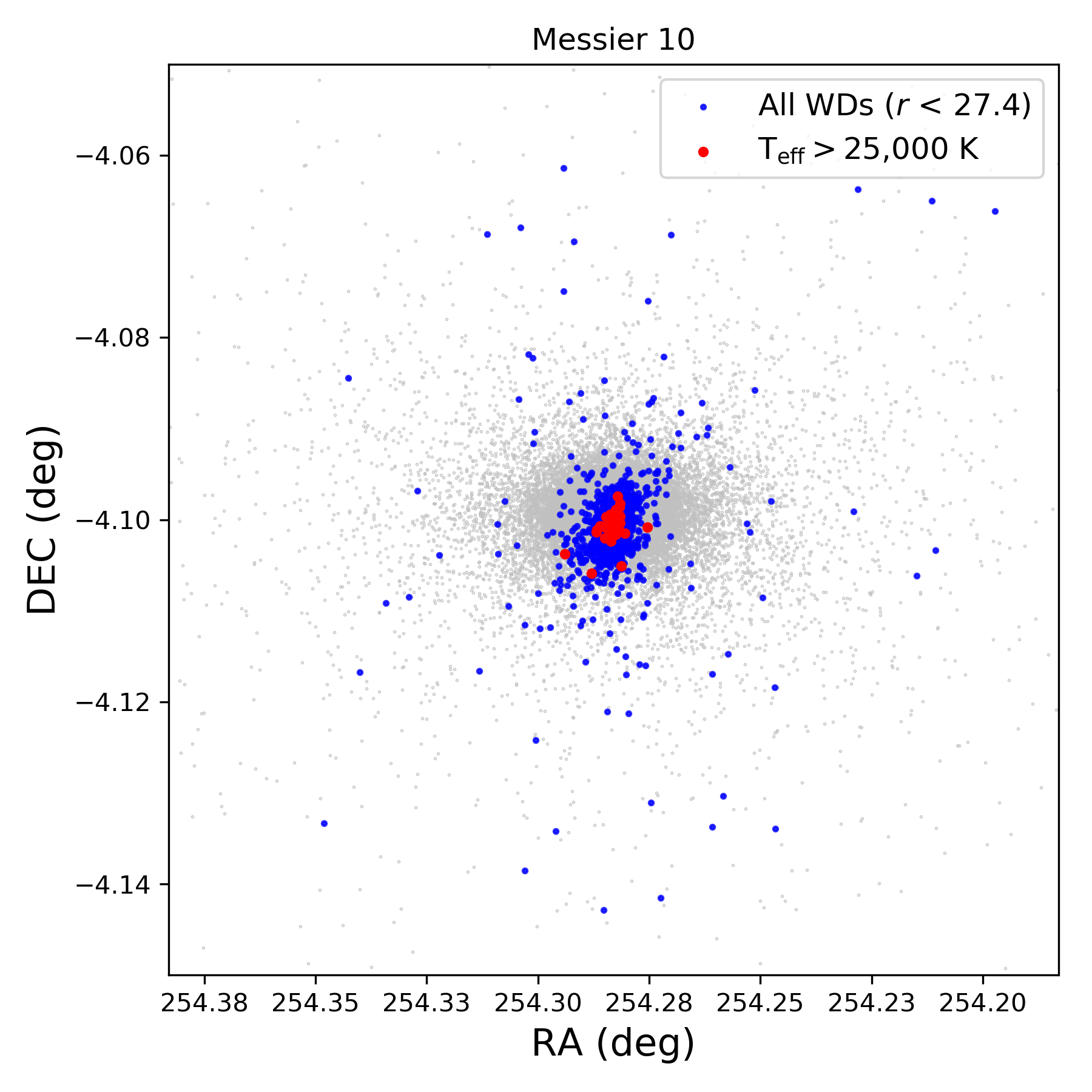}
	\includegraphics[angle=0,width=0.33\textwidth]{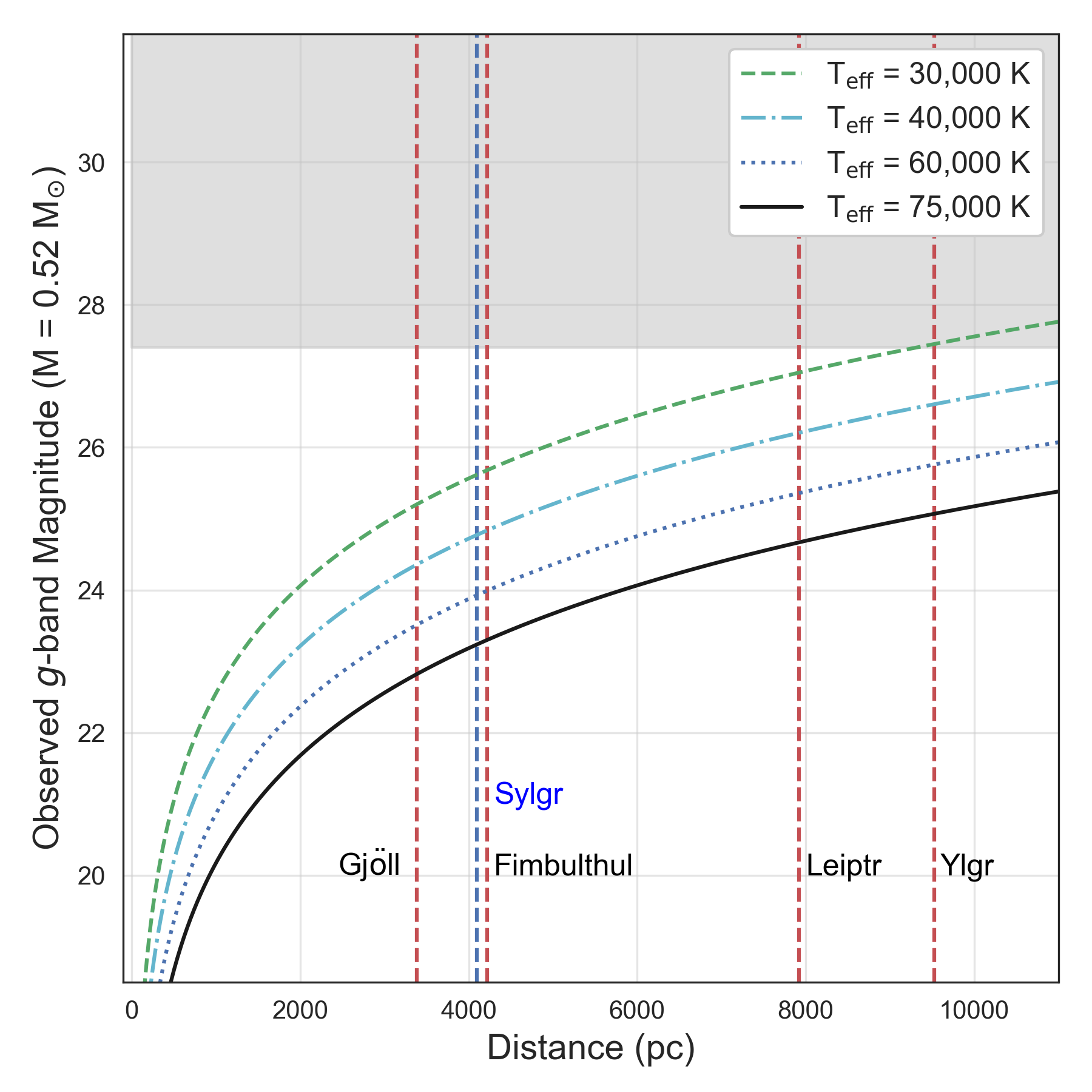}
	
	\caption{\textit{Left:} The observed $g$-band magnitude of a young white dwarf as a function of distance for four masses. The gray shaded region represents magnitudes beyond the 5$\sigma$ $g$-band magnitude limit of the LSST 10-year survey. Also marked are four globular clusters within the LSST footprint, of which one (NGC 288) also lies within the Roman Space Telescope HLS survey. \textit{Middle:} Simulation showing the white dwarf population in Messier 10, with objects hotter than 25,000\,K highlighted in red. \textit{Right:} The magnitude of young, hot, white dwarfs as a function of distance. Also marked are four nearby streams located within the LSST WFD footprint discovered by \cite{Ibata2019}.}
	\label{fig:clusters}
\end{figure*}

As the end stage of more than 97\% of stars, WDs are important tests of stellar evolution models. For a given initial (main-sequence) mass and age, their present-day masses contain information on the amount of mass lost during the progenitor's post-main-sequence evolution.

The relation between the initial mass and the resulting WD mass is called the initial-to-final mass relation (IFMR). While the IFMR does have theoretical prescriptions \citep[e.g,][]{Marigo2007, Choi2016}, it is usually measured semi-empirically from star clusters whose known age can be translated into an initial mass with an appropriate stellar isochrone. Thus, by measuring the masses of newly formed WDs in a cluster, one can relate the initial stellar masses to the final WD masses. By combining measurements over a wide range of ages (and hence initial masses), an empirical relation can be established between the two quantities \citep[see e.g,][]{Kalirai2008, Cummings2018}.

The combination of LSST and CASTOR will likely provide the best data for constraining the precise form of the IFMR. These surveys are sensitive to the hot, recently formed, WDs needed to use this technique, though ultimately, this relies on follow-up spectroscopy to measure masses. That is to say, these surveys will provide the imaging needed to identify hot WDs in distant clusters (up to 20 kpc as shown in Figure \ref{fig:distance}) which can be targeted for spectroscopy with large optical telescopes.

The left-hand panel of Figure~\ref{fig:clusters} shows the observed $g$-band magnitude for a young WD (T$_{\textrm{eff}}$ = 25,000\,K) as a function of distance for four different masses. For reference, the halo is currently producing WDs with masses close to 0.52\,M$_{\odot}$, while younger open clusters are producing WDs between 0.6\,M$_{\odot}$ and 1.0\,M$_{\odot}$ depending on their age. As Figure~\ref{fig:clusters} shows, many of the hot WDs in star clusters closer than $\sim$7.5\,kpc will be detected by LSST. 

We have used a modified version of our model to simulate the stellar population in the globular cluster Messier 10 (M\,10) to highlight the potential sample of high-temperature WDs which could be used for such a study. The resulting simulation can be seen in the middle panel of Figure \ref{fig:clusters}. We have set the mass of the cluster to be 2.25$\times10^{5}$\,M$_{\odot}$ with structural parameters as described in \cite{Gnedin1999}. The simulation predicts nearly 75,000 WDs present within M\,10, of which 7,500 are bright enough to be observed by the LSST 10-year survey (blue points). In this sample, we find that 63 objects have T$_{\mathrm{eff}} > $ 25,000 K (red points), with a few lying outside the core region. 

While crowding will play a role in the final number of objects discovered, our results suggest that there will be, at minimum, a handful of hot WDs observed in the outskirts of the cluster. Furthermore, since these objects are forming from low-mass stars, mass segregation may increase the number of progenitors in the outskirts of the cluster leaving a larger number of recently formed white dwarfs \citep{Spitzer1987, Parada2016b}. These photometric data, combined with upcoming large spectroscopic surveys like the proposed Maunakea Spectroscopic Explorer \citep{MSEBook2018, MSE} or the next generation Extremely Large Telescopes, will allow for the most accurate semi-empirical determination of the IFMR.

\begin{figure*}[!t]
	\includegraphics[angle=0,width=0.99\textwidth]{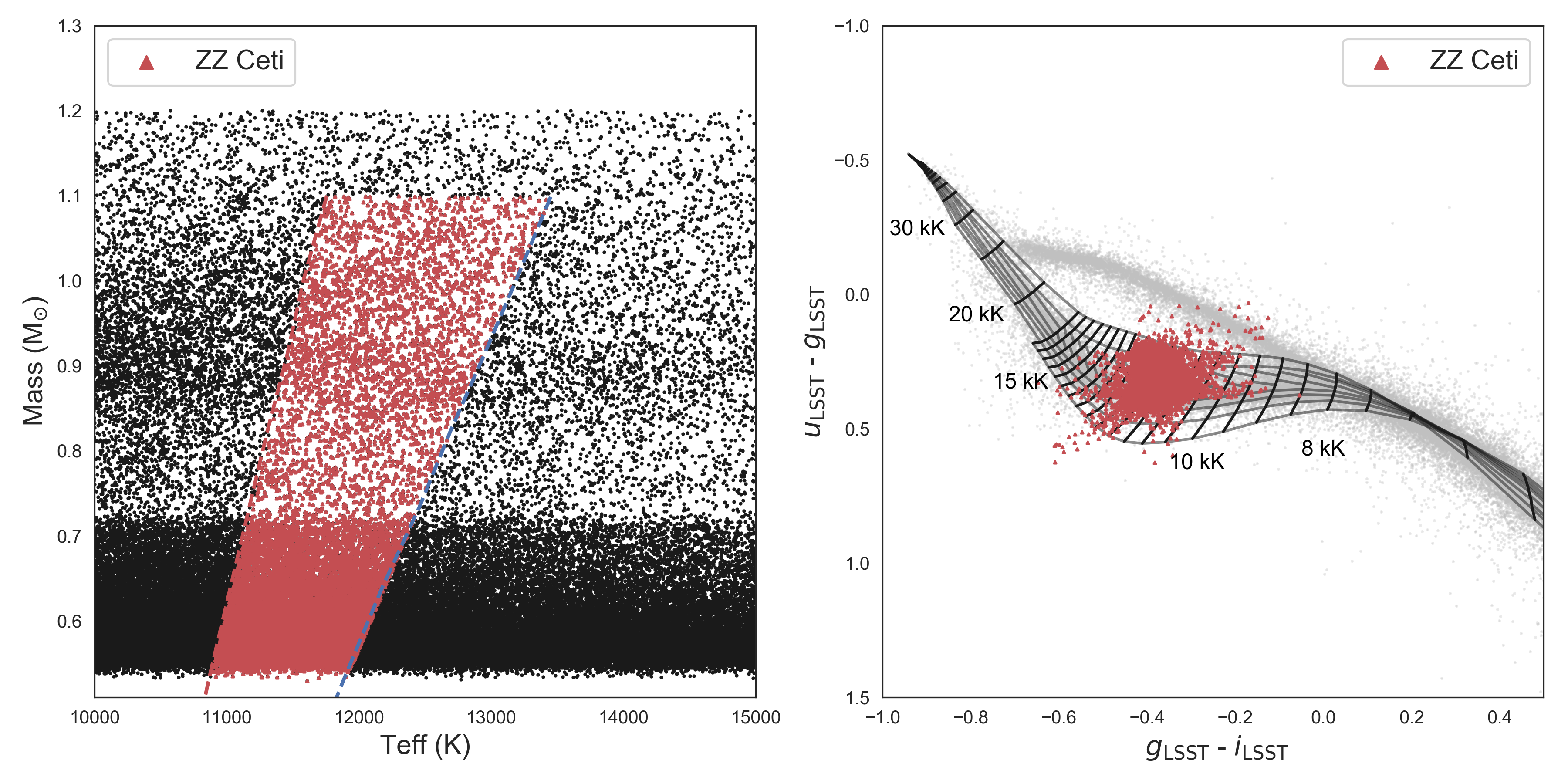}
	\caption{
	\textit{Left:} T$_{\rm eff}$--mass plot for white dwarfs in our LSST 5$\sigma$ proper motion sample (black points) highlighting the location of the ZZ Ceti instability strip defined by \cite{vangrootel2012} (red points).  \textit{Right:} The location of these objects in a color-color diagram, with solid black lines representing lines of constant T$_{\mathrm{eff}}$ and faded lines representing constant mass. Temperature values decrease with increasing ($g - i$) color as indicated. \bigskip }
	\label{fig:zzceti}
\end{figure*}

\subsection{Stellar Streams}

Recent studies have shown that the inner halo is crisscrossed with stellar streams, many of which are the tidal debris from halo globular clusters \citep{Malhan2018, Ibata2018, Ibata2019}. These streams were discovered using \textit{Gaia} DR2, relying mainly on the measured proper motions and photometry. LSST will provide equivalent information to much fainter magnitudes, suggesting that it will likely increase the number of known stellar streams.

With the addition of proper motions at fainter magnitudes, more stars within the known streams will be identifiable. To determine whether any WDs will be discovered within these streams, we select five nearby streams from \cite{Ibata2019} and simulate the magnitudes of a hot WD within each stream. These streams --- Gj\"{o}ll, Sylgr, Fimbulthul, Leiptr and Ylgr --- all lie within the LSST WFD footprint. Given that these streams are metal-poor, we assume that the stellar population is composed of stars similar to those in the oldest globular clusters (12\,Gyr) so their hot, young, WDs will be forming with masses of $\sim$0.52 M$_{\odot}$. We show the resulting magnitudes for these objects as a function of distance in the right-hand panel of Figure~\ref{fig:clusters}.

This exercise shows that LSST will provide $g$-band photometric measurements for the hot, young, WDs (T$_{\mathrm{eff}} > $ 30,000\,K) in all five streams, with the hottest objects (T$_{\mathrm{eff}} > $ 50,000\,K) in the nearest streams (Gj\"{o}ll, Sylgr and Fimbulthul) likely having reliable proper motions. 

The ability to reliably assign membership to these streams will rely on the measured properties of the stream, including the distance and proper motion, that will be determined from brighter more numerous stars. The addition of these WDs could provide information on the age if accurate masses can be measured. Because these streams can be closer to the Sun than many halo globular clusters, they could provide additional information on the low-mass end of the IFMR --- provided the stream age can be determined independently. 

\begin{figure*}[!t]
	\includegraphics[angle=0,width=0.49\textwidth]{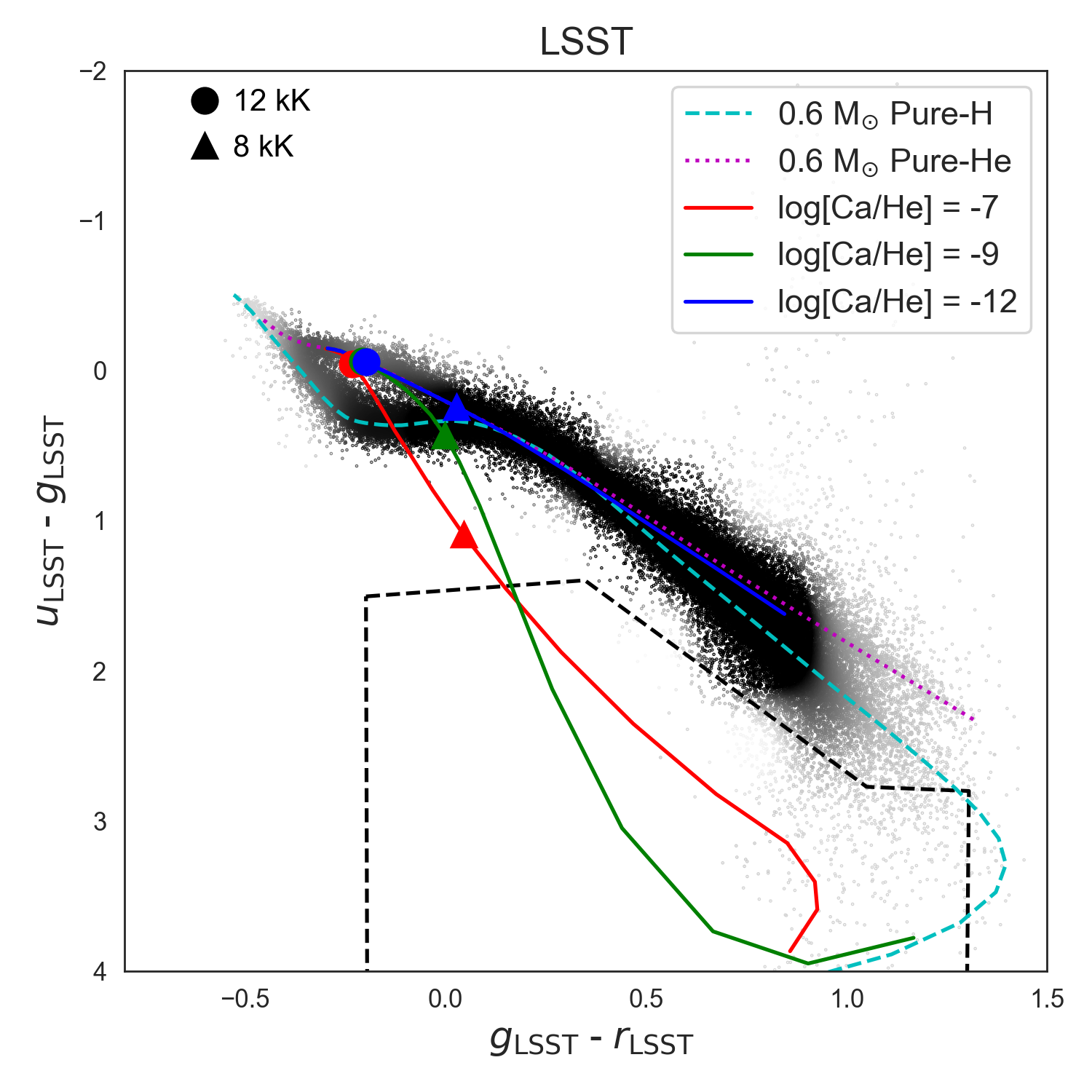}
	\includegraphics[angle=0,width=0.49\textwidth]{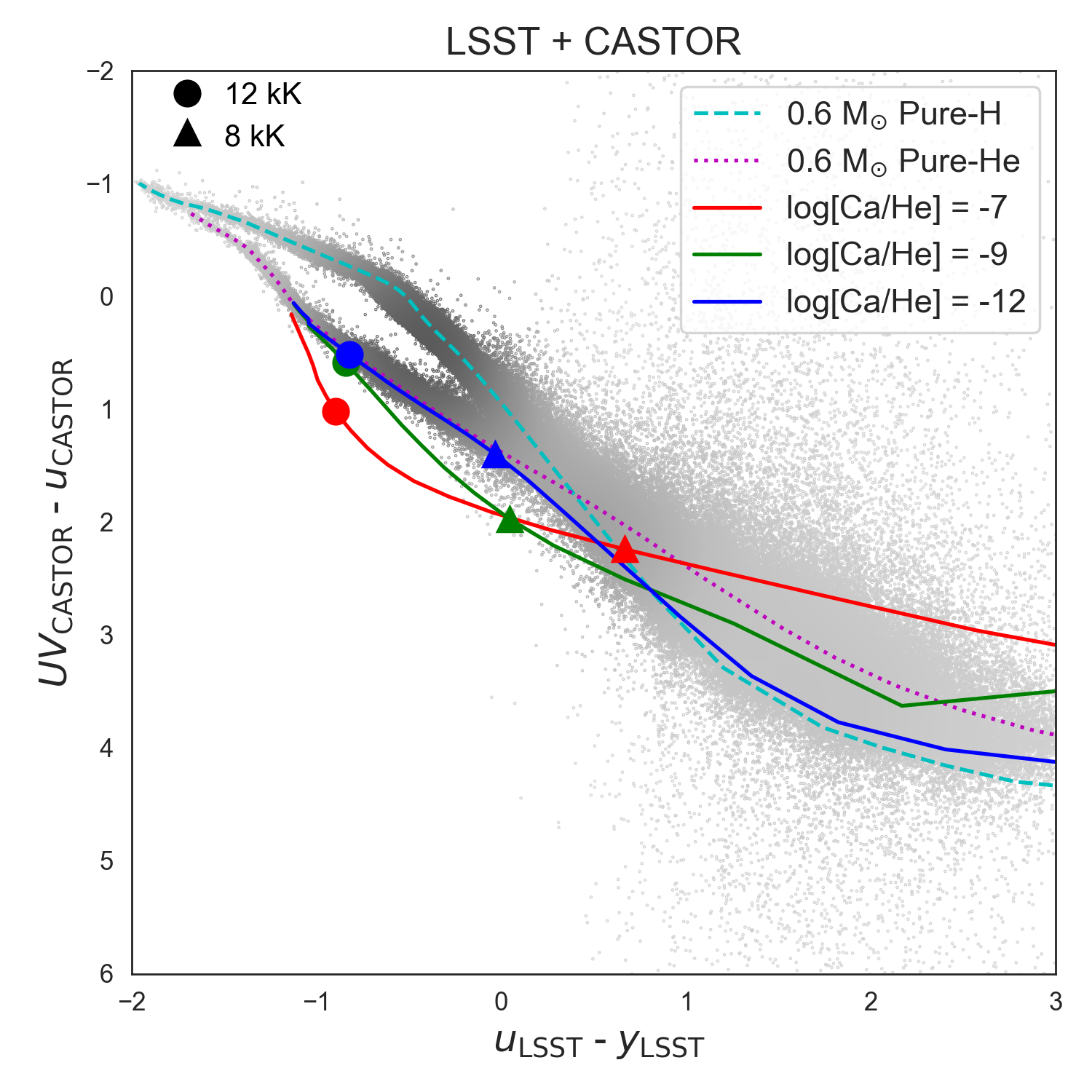}
	\caption{
	\textit{Left:} LSST color-color diagram showing the cooling tracks for pure-H, pure-He, and pure-He contaminated with Ca for which we use as a representation for a metal-polluted white dwarf. The black dashed box represents the color region used by \cite{Koester2011} and \cite{Hollands2017} to select cool metal-polluted white dwarfs. The circles on each plot represent T$_{\mathrm{eff}}$ = 12,000\,K, while the triangles represent T$_{\mathrm{eff}}$ = 8,000\,K.  \textit{Right:}  By combining LSST with other CASTOR it is possible to also separate higher T$_{\mathrm{eff}}$ metal-polluted white dwarfs, suggesting that a combination of colors can be used to select metal-polluted white dwarfs over a range of temperatures. \bigskip }
	\label{fig:DZ}
\end{figure*}

\subsection{Pulsating White Dwarfs}

One of the key science cases for time-domain surveys like the LSST is the study of luminosity fluctuations in stars. The first pulsating WD, HL Tau 76, was discovered by \cite{Landolt1968}, and subsequently, it was discovered that these objects had similar temperatures, leading to improved selection methods \citep{Fontaine1980}. The discovery of this new class of white dwarfs, called ZZ Cetis, was important as it allowed for the study of the mass of the stratified interior layers of the WD, while also allowing for an independent determination of the mass and atmospheric parameters \citep[see, e.g,][]{Kepler1995,Pech2006, Pech2006b,Castanheira2009,Althaus2010,Romero2012,Calcaferro2018}.

While there are different types of variable WDs, the vast majority of the currently known WD variables belong to the ZZ Ceti class. These objects have temperatures between 11,000 and 13,000\,K with pure-Hydrogen atmospheres, representing the coolest class of pulsating WDs \citep{vangrootel2012}. WDs with pure-Helium atmospheres also exhibit pulsations (and are classified as DBV), as do extremely low-mass WDs (ELMV and pre-ELMV). The remaining classes include much hotter WDs, many of which are still surrounded by their planetary nebula \citep{corsico2018}. An excellent comprehensive review of these classes and the current observational state of pulsating WDs can be found in \cite{corsico2019}. Here, we focus on the ZZ Ceti regime while noting that LSST will also uncover a large number of the other classes. In particular, LSST  should be able to accurately characterize the instability strips for the other classes of WD pulsators as their rarity requires this type of wide-field, multi-epoch survey.

ZZ Ceti stars experience pulsations as a result of the increased opacity in their outer atmosphere due to the recombination of hydrogen. This typically occurs between 11,000 and 13,000\,K, with this temperature range being a function of mass. \cite{Fontaine1982} suggested that all DA WDs will experience these pulsations (on the order of 100-1400\,s) as they cool and move across this instability strip, and observational evidence has since confirmed this hypothesis \citep{Fontaine2008}.

Therefore, to study the number of potential ZZ Ceti stars in the LSST, we use the observational boundaries from \cite{vangrootel2012} and apply them to our simulated sample of DA WDs with $>$5$\sigma$ proper motion objects. The boundary and results can be seen in Figure \ref{fig:zzceti}, where the left-hand panel shows the selection region, and the right-hand panel shows their location within a color-color diagram. We note that our model includes many massive objects above the specified mass regime of \cite{vangrootel2012}. Recent observations have uncovered a few examples of these massive pulsators \citep[1.20 $\pm$ 0.03\,M$_{\odot}$, see ][]{Hermes2013, Curd2017}, suggesting that the selection regime could be extended to the upper mass limit.

The total number of objects lying within this region is $\sim$200,000 in our 5$\sigma$ proper motion sample, with nearly 8,000 being brighter than the faintest ZZ Ceti discovered to date \citep[g = 19.5, see Table 5 in][]{corsico2019}.

Finally, we note that although we used the proper motion selected sample, ultimately the ability to study the pulsations will rely on the photometry and the ability to follow-up the observations at a higher cadence to more accurately determine periods. These objects will also benefit from follow-up spectroscopy to accurately determine their temperature and mass to define the boundaries of their respective instability strip.

\subsection{Metal Polluted White Dwarfs}

Our focus to this point has been on WDs with pure-Hydrogen or pure-Helium atmospheres as they comprise the vast majority of WDs. The atmospheres of white dwarfs, however, can also be polluted by metals. If the metal lines are the strongest spectral features the object is designated as spectral type DZ, although metal pollution can occur in other spectral types as well. For example, DA white dwarfs with metal lines present are designated as DAZ. The ability to detect such objects are of interest as the occurrence rates can reveal information regarding their formation mechanisms.

\begin{figure*}[!t]
	\includegraphics[angle=0,width=0.99\textwidth]{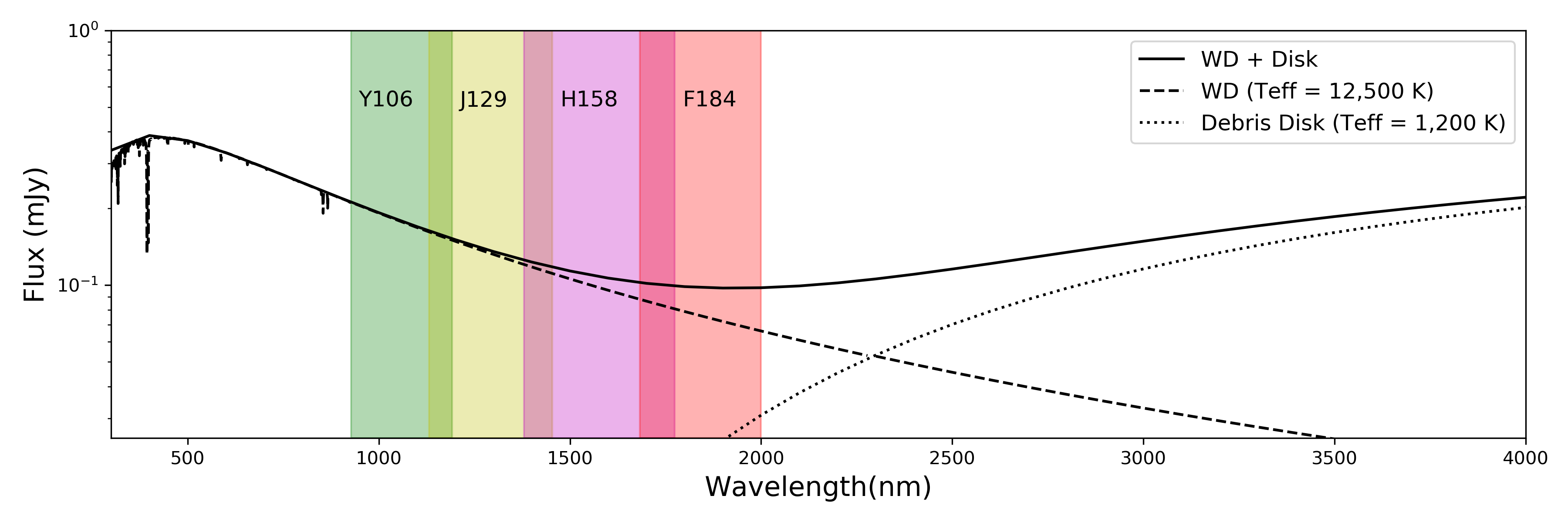}
	\caption{SED of a metal polluted WD (12,500\,K) with a warm debris disk (T = 1,200\,K) with the Roman Space Telescope filters highlighted. The presence of a debris disk will result in an excess of observed flux in the infrared, and will allow for the detection of such disks.}
	\label{fig:debris_disk}
\end{figure*}

Metal-polluted WDs can form as a result of episodic accretion from planetary material since any metals present in the photosphere at the time of formation would rapidly sink to the core \citep{Paquette1986}. This means that any metals present in the WD atmosphere can be used to constrain the bulk composition of the accreted bodies \citep{Zuckerman2007}, which has revealed a large amount of diversity in the planetary systems surrounding WDs \citep{Hollands2018}.

Such samples have typically relied on no more than a few hundred objects, however, the incidence of metal pollution in WDs has been found to be substantial. An example of this is the study of DA WDs with cooling ages between 20 and 200\,yr (T$_{\mathrm{eff}}$ = 22,000 - 27,000\,K) performed by \cite{Koester2014} using \textit{HST} UV spectroscopy. Their results suggest that at least 27\% of these objects show evidence of recent accretion, and the value may be closer to 50\%. LSST will inevitably observe large numbers of these objects and will be critical in identifying candidates for follow-up spectroscopy as many of the metal lines exhibited by these objects (Ca, Mg, Na) have strong absorption lines in the near-ultraviolet which results in decreased flux in the $u$-band. Indeed, the number of objects in our WFD 5$\sigma$ proper motion samples within the temperature range studied by \cite{Koester2014} is nearly 550,000.

\cite{Koester2011} and \cite{Hollands2017} used the SDSS photometry, in particular the $u$-band, to identify cool metal-polluted WDs in the SDSS. These objects separate from other WD types below $\sim$8,000\,K as the absorption from Ca I begins to dominate over the Balmer series within the $u$-band. To show this behavior, we have simulated metal-polluted WDs with three different calcium abundances in the LSST and CASTOR photometric bands as both surveys will contain a $u$-band. These models are adopted from those presented in \cite{Coutu2019}. The three different abundances represent the regimes between the most metal-polluted WDs ([Ca/He] = -7) to those with little pollution ([Ca/He] = -12). The resulting color-color diagram for LSST is presented in Figure \ref{fig:DZ}, with the selection regime of \cite{Koester2011} and \cite{Hollands2017} highlighted as the dashed box. 
As with the SDSS, LSST will allow for a separation of WDs in color-color diagrams as it has very similar passbands; however, the selection will be more accurate given the increased depth of the LSST and the ability to distinguish WDs from other point-sources via their proper motions.

We have also explored the possibility of selecting metal polluted WDs over a wider temperature range in the right-hand panel of Figure \ref{fig:DZ}. Here, we have cross-matched LSST with CASTOR in an attempt to isolate the warmer objects. Focusing on the [Ca/He] = -7 curve shows that the addition of a UV filter and a broad wavelength coverage can indeed separate hot metal-polluted objects with a large amount of calcium absorption at 12,000\,K. This suggests that a combination of colors can be used to identify a larger sample of metal-polluted WDs for follow-up observations with new and upcoming spectroscopic instruments, the results of which will allow for the study of the rate of incidence and physical processes at work in such systems.

\subsection{Debris Disks and Sub-stellar Companions}

The metals present in WDs are thought to originate from the accretion of either debris disks or remnant planetesimals \citep[see e.g,][]{Kilic2006, Farihi2010, Koester2014}. The debris disks are warm, with temperatures ranging from a few hundred to a few thousand Kelvin and occur around 1 - 5\% of WDs \citep{Barber2012}, although the fraction of systems showing evidence for recent accretion may be as high as 50\% \citep{Koester2014}.

These debris disks are typically discovered by comparing model SEDs of the WD using properties determined through optical spectroscopy to the observed infrared magnitudes, where any excess infrared emission is attributed to the debris disk. We have highlighted this point in Figure \ref{fig:debris_disk}, where we have shown the SED of a metal-polluted WD with a surface temperature of 12,500\,K and surrounded it by a debris disk of 1,200\,K. This disk is modeled after the G29-38 \citep{Zuckerman1987, Reach2005} disk presented in \cite{Jura2003} (see equation 3). This disk is modelled as a face-on, flat, opaque ring surrounding the host star with an inner temperature of 1,200\,K at 0.14 R$_{\odot}$. We set the distance to be 100\,pc from the Sun.

The debris disk is completely invisible over the optical wavelengths covered by the LSST, although the excess emission from the debris disk becomes apparent in the F184 bands from the Roman Space Telescope, with a difference in observed magnitude of 0.27 AB mag. This difference should be detectable to a limit of $\sim$25th magnitude in the F184 band. This shows that the combination of optical data from the LSST, combined with the NIR photometry from the Roman Space Telescope or Euclid, will be ideal for detecting such objects. 

To estimate the number of WDs with debris disks that might be present within the HLS, we apply the selection method from \cite{Wilson2019}, T$_{\mathrm{eff}} = 14,000 - 31,000$, and apply their resulting rate of debris disks which show an infrared excess, 1.5$^{+1.5}_{-0.5}$\%, to our sample of LSST 5$\sigma$ proper motions combined with the Roman Space Telescope sample. Our results suggest that this method could detect on the order of 355 WDs with debris disks in the HLS footprint. 

We note, however, a few caveats that come with this number. First, the disk modeled above is a more extreme example of debris disks surrounding white dwarfs as its inclination is thought to be close to zero. \cite{rocchetto2015} showed that such bright, face-on disks comprise less than 10\% of all known white dwarf debris disks, while most have luminosities that are less than 20\% of such a model. The remaining disks, as well as those with lower temperatures and/or at larger distances, will be more difficult to detect. As an example, changing the inclination of the G29-38 disk from face-on to an inclination of 75$^{\circ}$ decreases the flux excess in the F184 band to 0.07 AB mag, and decreasing the temperature, size, or opacity will reduce this value even further. Applying the results from \cite{rocchetto2015} still suggests that a few dozen bright, dusty, debris disks surrounding white dwarfs with an excess in the F184-band will be detectable in the HLS.

Another caveat is that debris disks have been found in cooler WDs \citep{Kilic2006}, and as such a larger number of WDs with debris disks may await discovery. Ultimately, the combination of LSST and Euclid and/or the Roman Space Telescope will help in determining the fraction of WDs with 2$\mu$m emission over a wide range of temperatures.

\section{Summary}
\label{sec:summary}

In this paper, we have used a new model to simulate the WD populations that will be detected in three pending and one proposed wide-field surveys (LSST, Euclid, the Roman Space Telescope and CASTOR) whose depth, wide fields and broad wavelength coverage across the UV/optical/NIR region will open new avenues in WD research. Our key findings are as follows:

\begin{itemize}
    \item We predict that LSST, at its final 10-year depth, will detect more than 150 million WDs in its WFD survey, although identifying many of these white dwarfs could be a challenge, depending on the availability of parallaxes and/or proper motions, the precision of those measurement, and availability of supplemental data from other facilities.
    
    \item We identify three distinct magnitude regimes in which different parameters measured by LSST can be used to optimally select WDs. At $r \lesssim$ 20.0, we predict $\sim$ 300,000 WDs can be identified using parallaxes measured to a precision of 5$\sigma$ or better. In the range $20.0 \lesssim r \lesssim 23.5$, we expect $\sim$ 7 million WDs can be identified using proper motions measured to a precision of  5$\sigma$ or better. We anticipate the impact of LSST for WD research will be highest in this regime owing to the tremendous gain in sample size and the generally low contamination rate. Fainter than $r \sim 23.5$, WDs will need to be selected from photometry alone. This will be particularly difficult contamination from  other  stellar  contaminants  and  inter-loping background galaxies becomes substantial at faint magnitudes. The expanded wavelength coverage and excellent image quality provided by the space-based facilities will be especially important for identifying clean WD samples at these faint magnitudes.  

    \item Proper motions will continue to be a key tool for identifying WDs within upcoming surveys. While LSST, and potentially the Roman Space Telescope observations acquired for its High Latitude Survey, will yield accurate proper motion measurements, the cadence adopted by other facilities or programs will be important for maximizing WD yields.
    
    \item The Roman Space Telescope High Latitude Survey represents the largest overlapping region between the four surveys. This region will benefit from excellent star-galaxy classification and simultaneous UV, optical and NIR photometry. Several WD science cases, including from initial-to-final mass relation and the frequency of WD debris disks and sub-stellar companions --- will be enabled by the multi-wavelength observations in this region.
    
    \item With simulations, we illustrate how the WD luminosity function --- for both the disk {\it and} the halo --- will be well constrained beyond the turn-off point, leaving only systematic uncertainties for all but the most extreme ends.
    
    \item UV observations will be particularly useful for identifying the hottest and youngest WDs in star clusters out to 20\,kpc. Simulations for the globular cluster Messier 10 show that a handful of young, hot, WDs will be observed with the LSST. These objects will need spectroscopic follow-up with upcoming large telescopes to determine an accurate mass to derive an empirical initial-to-final mass relation.
    
    \item LSST will observe more than 200,000 ZZ Cetis, providing $\sim$ 800 independent flux measurements over the 10-year survey lifetime; we estimate that more than 8000 of these objects will be brighter than $g$= 19.5. In addition, the survey will help constrain the instability strip boundaries for other types of pulsating WDs.
    
    \item LSST will also observe WDs with polluted atmospheres. We show that a combination of the UV and optical surveys will be useful for identifying WDs with metal-polluted atmospheres over a wider range of temperatures when compared to data solely acquired as part of the LSST.

    \item NIR photometry is critical for the discovery of debris disks surrounding WDs. Using current estimates for the fraction of WDs having associated debris disks, we estimate that combined LSST and Roman Space Telescope observations will uncover dozens of white dwarfs with debris disks. 
    
\end{itemize}

This study highlights only a few of the many exciting science cases involving WDs during the coming decade, including the WD luminosity function of the Milky Way's components, the initial-to-final mass relation, and the variable and pulsating WDs, to name a few. LSST and other surveys will provide a multitude of science cases beyond what has been presented here, and examples in the literature include the discovery of companion planets \cite[see e.g,][]{lund2018,cortes2019} and the study of WDs through gravitational waves \cite{korol2019}. The sheer number of objects to be revealed in these surveys will inevitably lead to unforeseen discoveries, pointing the way to a decade of exciting WD research.

\acknowledgements

This work was supported in part by the Natural Sciences and Engineering Research Council of Canada (NSERC). We would like to thank Pierre Bergeron and Patrick Dufour for providing their white dwarf cooling models in the relevant passbands.

This material is based upon work supported in part by the National Science Foundation through Cooperative Agreement 1258333 managed by the Association of Universities for Research in Astronomy (AURA), and the Department of Energy under Contract No. DE-AC02-76SF00515 with the SLAC National Accelerator Laboratory. Additional funding for the Rubin Observatory comes from private donations, grants to universities, and in-kind support from LSSTC Institutional Members.

\end{document}